\documentclass[a4paper]{jpconf}
\usepackage{graphicx}
\usepackage{eurosym}
\usepackage{amssymb}
\usepackage{braket}
\usepackage{amsfonts,cite}
\usepackage{amsmath,epsfig}
\usepackage{multirow}
\usepackage{subcaption}
\usepackage{tikz}
\usepackage{float}
\usepackage{xcolor}
\newbox\mybox

\newcommand\fverb{\setbox\mybox=\hbox\bgroup\verb}
\newcommand\fverbdo{\egroup\medskip\noindent\fbox{\unhbox\mybox}\ }
\newcommand\fverbit{\egroup\item[\fbox{\unhbox\mybox}]}
\DeclareMathOperator{\arccosh}{arccosh}
\DeclareMathOperator{\arcsinh}{arcsinh}

\input{tcilatex}
\begin{document}
\title{An introduction to ${\cal PT}$-symmetric quantum mechanics - time-dependent systems }
\author{Andreas Fring} 	
\address{
Department of Mathematics, City, University of London,
Northampton Square, \\ London EC1V 0HB, UK }
\ead{a.fring@city.ac.uk}

\begin{abstract}
	I will provide a pedagogical introduction to non-Hermitian quantum 
	systems that are ${\cal PT}$-symmetric, that is they are left invariant under a simultaneous parity transformation (${\cal P}$) and time-reversal (${\cal T}$). I will explain how generalised
	versions of this antilinear symmetry can be utilised to explain that these type of systems
	possess real eigenvalue spectra in parts of their parameter spaces and how to set up
	a consistent quantum mechanical framework for them that enables a unitary time-evolution.
	In the second part I will explain how to extend this framework to explicitly time-dependent Hamiltonian systems and report in particular on recent progress made
	in this context. I will explain how to construct the essential key quantity in this
	framework, the time-dependent Dyson map and metric and solutions to the time-dependent Schr\"odinger equation, in an algebraic fashion, using time-dependent Darboux transformations, utilising Lewis-Riesenfeld invariants, point
	transformations and some approximation methods. I comment on the ambiguities
	of this metric and demonstrate that this can even lead to infinite series of metric
	operators. I conclude with some applications to ${\cal PT}$-symmetrically coupled oscillators, demonstrate the equivalence of the time-dependent double wells and unstable
	anharmonic oscillators and show how the unphysical ${\cal PT}$-symmetrically broken regions in the parameter space for the time-independent theory becomes physical in
	the explicitly time-dependent systems. I discuss how this leads to a prolongation of the
	otherwise rapidly decaying von Neumann entropy. The so-called sudden death of the entropy is stopped at a finite value.\footnote{These notes are based on lectures presented at the International Conference on Quantum Phenomena,
		Quantum Control and Quantum Optics, virtually held at Cinvestav Mexico, 25/10-29/10 2021.}
\end{abstract}

\section{Introduction}
In principle it has been known for over sixty years that non-Hermitian Hamiltonians which commute with antilinear operators may possess real eigenvalue spectra \cite{EW} and may also be associated to a consistent quantum mechanical framework \cite{Dyson}. While scepticism on the practicability of treating such systems was expressed in the early publications, e.g. footnote 9 in  \cite{Dyson}, more systematic investigations of the concrete details involved and examples were only worked out much later \cite{Urubu}. Most notably the concrete study of simple ${\cal PT}$-symmetric potentials in \cite{BB} led to a wider interest in the subject. Meanwhile there exists a wide general consensus on the conceptual consistency and feasibility to solve these type of systems, especially when they are autonomous \cite{Rev2,Alirev,bagarello2015non,PTbook}. It is also well-known that there are occasionally mathematical issues related to the boundedness of some of the operators involved \cite{siegl2012metric,bagarello2013self,bagarello2017pseudo}, but the multitude of physical applications and experimental verifications leave no doubt on the viability of the general theoretical framework. Several special issues, mostly centred around annual conferences, have been assembled, e.g. \cite{special1,special2}. For recent results and applications in diverse physical systems see for instance the issue \cite{PTseminar} based around a virtual seminar series \cite{virtual}.

For explicitly time-dependent systems the situation is still less complete. The first treatments of ${\cal PT}$-symmetric time-dependent non-Hermitian Hamiltonian systems were carried out in \cite{CA,CArev}. Since then there have been a number of publications on the subject \cite{time1,time2,time3,time4,time5,time6,mehri2008geometric,znojil2009three,BilaAd,gong2010geometric,time7,maamache2017pseudo,khantoul2017invariant,zhang2019time,mostafazadeh2020time} including some in which the conceptual consistency of time-dependent non-Hermitian Hamiltonian systems has been put into question and some with disputable starting points. In most cases the controversies were simple rooted in linguistic differences and conceptual inconsistencies of definitions, but at times precise abstractions have not been made. As most the controversial issues seem to be settled by now, we will not revisit the debates and simply draw here on those publications when we have agreement with the following recent body of work \cite{fringmoussa,fringmoussa2,AndTom1,AndTom2,AndTom3,AndTom4,AndTom5,cen2019time,fring2019eternal,fring2020time,frith2020exotic,frith2019time,juliathesis,BeckyAnd1,BeckyAnd2,fring2021perturb,fring2021exactly,fring2021infinite,beckythesis} that constitutes the main basis of this presentation.  We stress that our interest here is on self-consistent non-Hermitian ${\cal PT}$-symmetric systems that are distinct from  dissipative systems \cite{moiseyev2011non,Gilary2005calc}, which in contrast have to be open.

These notes are organised as follows: In section 2 we provide an introduction to non-Hermitian time-independent systems. We recall some prominent examples of non-Hermitian non-dissipative systems from the literature, provide various explanations for the reality of their spectra by using generalised versions ${\cal PT}$-symmetry, pseudo/quasi-Hermiticity and Darboux transformations. We explain in general and with some worked out examples how to formulate a consistent quantum mechanical framework for them by constructing Dyson maps and introducing a new metric for inner products. In section 3 we introduce explicitly time-dependent Hamiltonian systems by discussion the key concepts and provide a detailed discussion on various possibilities on how to solve these systems. In section 4 we discuss some application of time-dependent non-Hermitian Hamiltonian systems, in particular focussing on the different types of behaviour of the von Neumann entropy in three distinct ${\cal PT}$-symmetric regimes.  
    
	\section{${\cal{PT}}$-symmetric quantum mechanics - time-independent $H$}
	\subsection{Hermiticity is only a sufficient but not a necessary requirement}
	Our key objective is here to provide meaningful interpretation to quantum mechanical systems based on non-Hermitian Hamiltonians when compared to the most common standard approaches in which one assumes the Hamiltonians to be Hermitian. Let us therefore first recall the principal reasons for why Hermiticity is a very useful property to have in a physical system before arguing that the same features can be achieved when starting from non-Hermitian Hamiltonian systems. 
	
	The first reason is the fact that Hermiticity guarantees the reality of energies. This is easily seen when starting from the time-independent Schr\"odinger equation (TDSE) for some state vector $\vert \psi \rangle$ involving a time-independent Hamiltonian $H$ and its conjugate $H^\dagger$
	\begin{equation}
		   H \vert \psi \rangle     =E \vert \psi \rangle \qquad \text{and} \qquad
		   \langle \psi \vert H^\dagger = E^*  \langle \psi \vert . \label{TISE}
	\end{equation}	
Multiplying the first equation by the bra state $ \langle \psi \vert$ from the left and the second equation by the ket state $\vert \psi \rangle $ from the right, and subsequently taking the difference we obtain 
	\begin{equation}
 	\langle \psi \vert H \vert \psi \rangle - \langle \psi \vert H^\dagger \vert \psi \rangle    = (E- E^*)   \langle \psi \vert \psi \rangle . \label{realEV}
\end{equation}	
Thus when $H$ is Hermitian, i.e. $ H= H^\dagger$, the left hand side vanishes and since $\langle \psi \vert \psi \rangle \neq 0$ it follows that the energy $E$ must be real.   

The second reason is the fact that Hermiticity ensures the conservation of probability densities. This is easily seen by starting from the evolution of a state at time $t=0$ to a state at time $t$ 
\begin{equation}
\left\vert \psi (t)\right\rangle =e^{-iHt/\hbar}\left\vert \psi (0)\right\rangle .
\end{equation}
Sometimes we will set Planck's constant $\hbar=1$. Taking the conjugate of this equation, multiplying by $\left\langle \psi (t)\right\vert$ from the left and assuming once more that the Hamiltonian to be Hermitian we obtain 
\begin{equation}
\left\langle \psi (t)\right\vert \left. \psi (t)\right\rangle =\left\langle
\psi (0)\right\vert e^{iH^{\dagger }t/\hbar}e^{-iHt/\hbar}\left\vert \psi
(0)\right\rangle =\left\langle \psi (0)\right\vert \left. \psi
(0)\right\rangle .
\end{equation}
This means the probability density at the time $t=0$ is the same as at any other arbitrary time $t$, i.e. it is conserved. Therefore when considering non-Hermitian systems it seems natural to think of dissipative systems that do not preserve probabilities. However, these type of systems are open systems requiring an environment to be embedded into \cite{RotterRev,friedrich2006theoretical} and do not allow for a self-consistent description. Here we will focus on systems that do allow for a self-consistent formulation, leaving aside the much more general question of in as much as quantum mechanics can be regarded as complete.  

As we shall argue in detail below, these two properties can also be obtained for theories based on non-Hermitian Hamiltonians, although in a less obvious way. It was already pointed out by Wigner more than sixty years ago \cite{EW} that {\it  Operators $\mathcal{O}$ that are left invariant with respect to an antilinear operator $\mathcal{I}$, i.e. $\left[ \mathcal{O},\mathcal{I}\right] =0$, and whose eigenstates $ \left\vert \Phi \right\rangle $ also respect this symmetry, $\mathcal{I} \left\vert\Phi \right\rangle= \left\vert \Phi \right\rangle$, have real eigenvalue spectra.} Moreover a consistent quantum mechanical framework can be provided by defining a new metric \cite{Urubu,BB,Mostafazadeh:2001nr,PTbook}. We will make these statements more precise in what follows.

	\subsection{Seminal and pre-historic examples in the literature}
	Motivated by a variety of reasons, theories based on non-Hermitian Hamiltonians have been considered over the years in different kinds of contexts, such as for instance in an attempt to describe high energy hadron scattering in form of lattice Reggeon theories \cite{cardy1975reggeon}, as discrete quantum spin chain versions of non-unitary conformal field theories \cite{gehlen1}, as quantum field theories \cite{Holl} or as effective theories obtained from string theory \cite{das2007s}. From a current perspective one may question whether these systems have been treated adequately, but these selected examples illustrate that one is often naturally led to non-Hermitian systems in almost all subfields of physics. 
	
	 \begin{figure}[h]
		\noindent	\begin{minipage}[b]{0.48\textwidth}     \!\!\!\! \!\!\!\! \includegraphics[width=\textwidth]{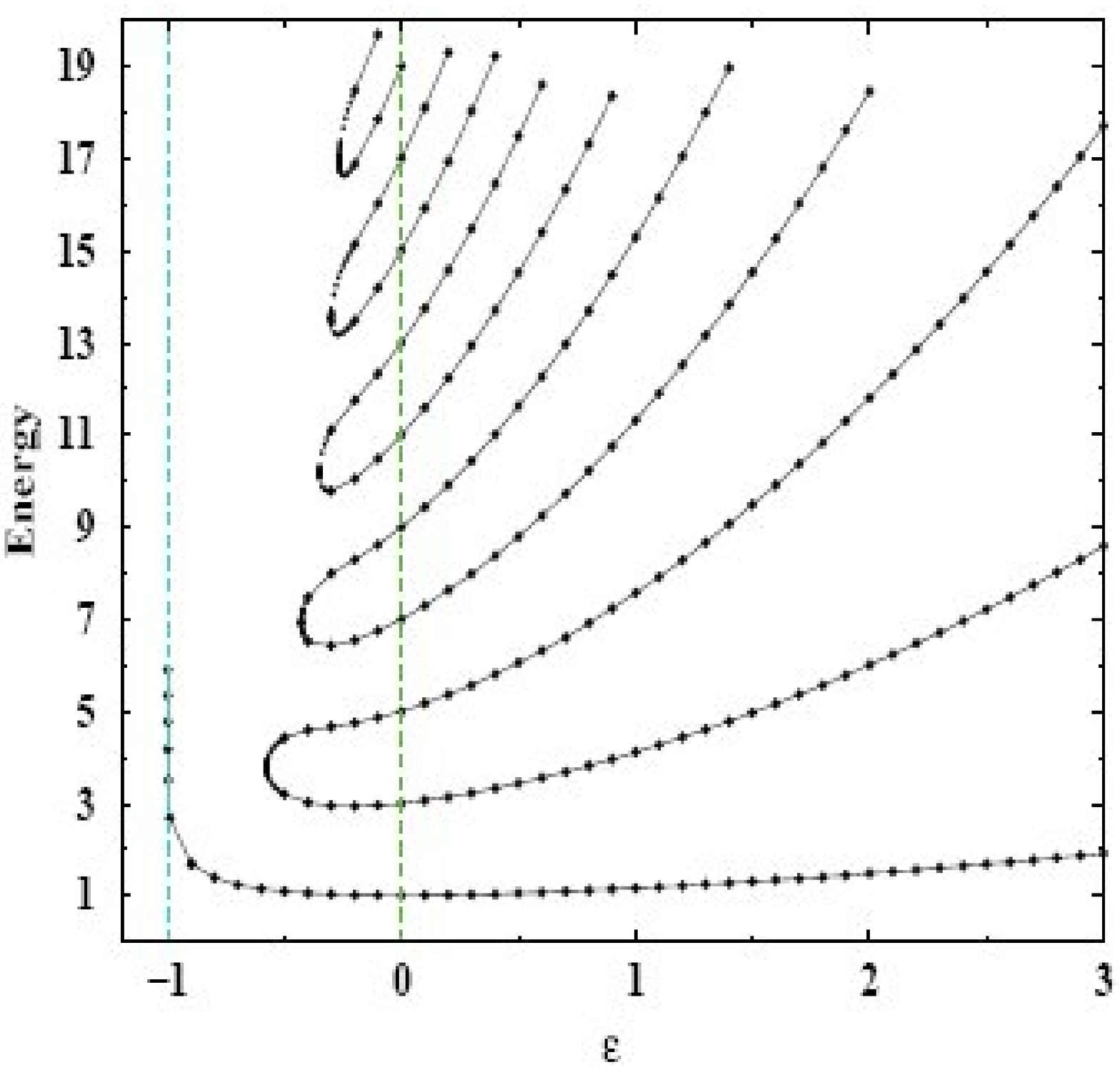}
		\end{minipage}
		\begin{minipage}[b]{0.52\textwidth}      \includegraphics[width=\textwidth]{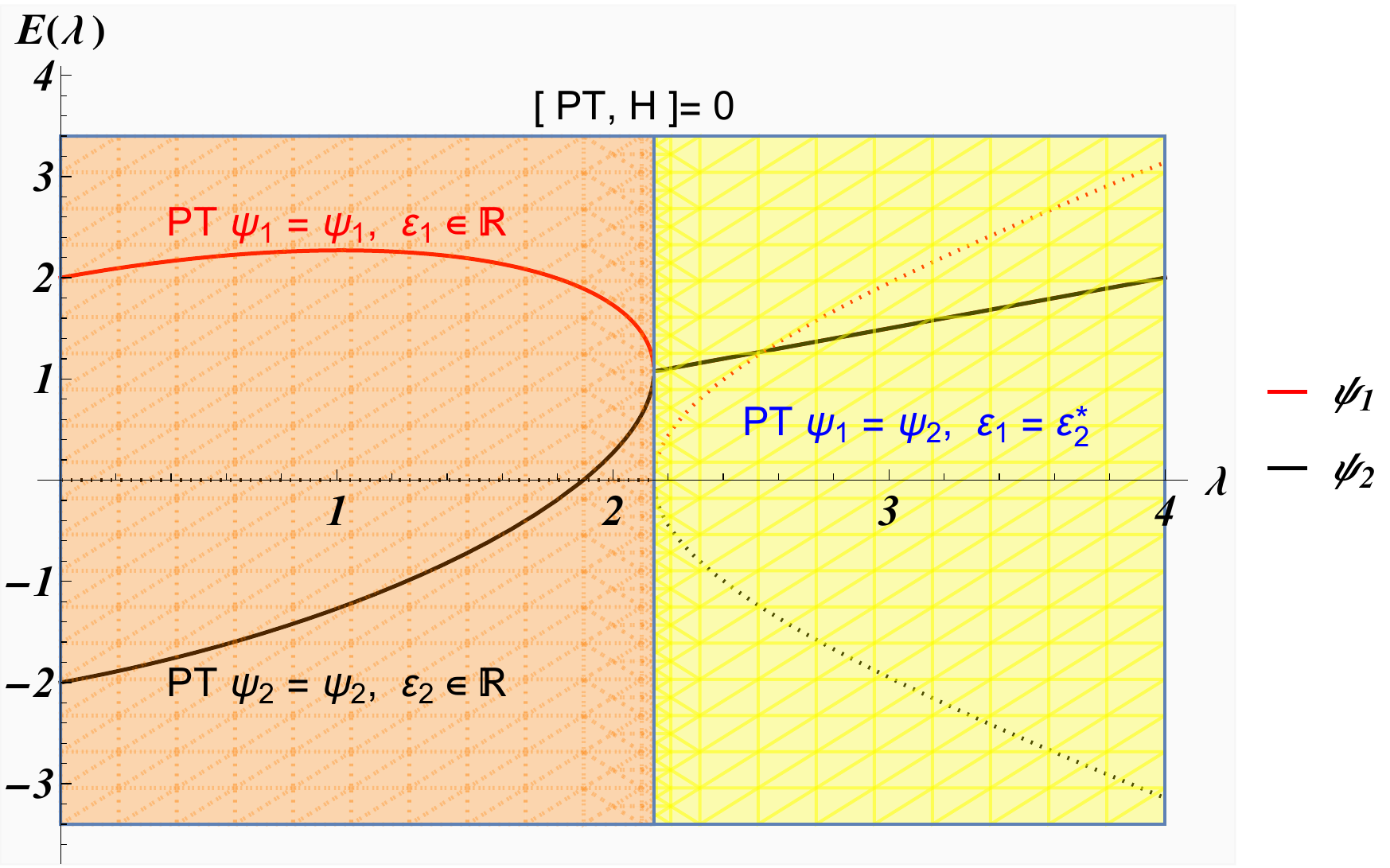}    
		\end{minipage}
		\caption{Panel (a): Positive, discrete and real energy spectrum for the non-Hermitian Hamiltonian $H_{BB}$ in (\ref{HBB}) as a function of $\varepsilon$. (Figure taken from \cite{BB}.) Panel (b): Level crossing in a two-level energy spectrum as a function of a continuous system parameter $\lambda$ in $\mathcal{PT}$-symmetric (orange) versus spontaneously broken $\mathcal{PT}$ (yellow) regimes. Real parts correspond to solid and imaginary parts to dotted lines.  }
		\label{figBB}
	\end{figure}   
	The first general considerations to treat non-Hermitian systems as self-consistent physical descriptions were carried out in \cite{Urubu}, but a wider community was only attracted several years later in 1998 by the observations made in the seminal paper by Bender and Boettcher \cite{BB} who studied the energy spectrum of the non-Hermitian Hamiltonian system
	\begin{equation}
	H_{BB}=\frac{1}{2}p^{2}+x^{2}(ix)^{\varepsilon },\qquad \text{for }\varepsilon \in \mathbb{R}. \label{HBB}
	\end{equation}
	One may view $H_{BB}$ as a deformation of the harmonic oscillator corresponding to the value $\varepsilon = 0$. The remarkable feature of its spectrum, depicted in figure \ref{figBB} panel (a), is that despite the fact of the Hamiltonian being non-Hermitian it is real for $\varepsilon \geq 0$.  In the region $-1 < \varepsilon < 0$ only a finite number of real eigenvalues survive. Noteworthy is also the special case $\varepsilon = 2$ corresponding to the potential $V=-x^4$. Even though this potential is real, one would still expect the theory to be ill-defined as the potential is unbounded from below. We return to this point below in section \ref{anharmsec} and explain why this is not necessarily the case.

 \subsection{Non-Hermitian models from noncommutative space-time structure}
 In fact one does not have to start from a concrete model to be led to non-Hermitian Hamiltonian systems, but instead consider Hamiltonian systems on certain types of deformed space-time structures. Depending on the representation of the underlying canonical commutation relation {\bf any} model on such a space becomes non-Hermitian. Most interesting are those deformed spaces that lead to generalized versions of the Heisenberg uncertainty (GUP) relations \cite{Kempf1,Kempf2}. To see this we follow \cite{AFBB} and start from a very generic $q$-deformation of Heisenberg's canonical commutation relations obeyed by creation and annihilation operators $a^\dagger$ and $a$
 \begin{equation}
 	aa^{\dagger }-q^{2}a^{\dagger }a=q^{g(N)},
 	\label{aa}
 \end{equation}
where $g(N)$ is an arbitrary function of the number operator $N=a^{\dagger }a$ and $q$ is a deformation parameter. Introducing the variables $X$ and $P$ and expressing them linearly in terms of $a^\dagger$, $a$
 \begin{equation}
 	X=\alpha a^{\dagger }+\beta a,~~~P=i\gamma a^{\dagger }-i\delta a,\qquad ~%
 	\text{ }\alpha ,\beta ,\gamma ,\delta \in \mathbb{R}
 \end{equation}
we compute their mutual commutation relations with the help of (\ref{aa})
 \begin{equation}
 	  \left[ X,P\right] =i\hbar q^{g(N)}(\alpha
 	\delta +\beta \gamma ) +\frac{i\hbar (q^{2}-1)}{\alpha \delta +\beta \gamma }\left( \delta \gamma
 	X^{2}+\alpha \beta ~P^{2}+i\alpha \delta XP-i\beta \gamma PX\right) .
 \end{equation}
 To simplify matters we carry out the nontrivial limit $\beta \rightarrow \alpha$, $\delta \rightarrow \gamma$, $g(N) \rightarrow 0$,
 $q \rightarrow e^{2 \tau \gamma^2}$, $\gamma \rightarrow 0$ obtaining
 \begin{equation}
 	\left[ X,P\right] =i\hbar \left( 1+\tau P^{2}\right) .  \label{12}
 \end{equation} 
We observe here that in the limit of vanishing deformation parameter $\tau \rightarrow 0$ relation (\ref{12}) simply reduces to the standard canonical commutation relation $[x_{0},p_{0}]=i\hbar$ satisfied by the coordinate $x_0$ and the momentum operator $p_0$ with $X \rightarrow x_0$ and $P \rightarrow p_0$. A natural representation for $X,P$ in terms of $x_{0},p_{0}$ is easily guessed 
\begin{equation}
 X=(1+\tau p_{0}^{2})x_{0}, \quad \text{and} \qquad P=p_{0}.  \label{repXP}
\end{equation}
Thus, with the standard inner product the operator $X$ is not Hermitian
\begin{equation}
X^{\dagger }=X+2\tau i\hbar P~ \quad \text{and} \quad P^{\dagger }=P .
\end{equation}
This means that in general {\bf any} Hamiltonian that would be Hermitian when expressed in terms of the standard coordinate and momentum  $H(x_0,p_0)$ becomes non-Hermitian on a noncommutative space with the replacements $x_0 \rightarrow X$, $p_0 \rightarrow P$. This is easily seen for the simple example of the harmonic oscillator on such a space when using the representation (\ref{repXP})
\begin{eqnarray}
	H_{ho}(X,P) \!\! &=& \!\! \frac{P^{2}}{2m}+\frac{m\omega ^{2}}{2}X^{2}  \label{3} \\
\!\! &=& \!\! \frac{p_{0}^{2}}{2m}+\frac{m\omega ^{2}}{2}(1+\tau p_{0}^{2})x_{0}(1+\tau
	p_{0}^{2})x_{0} \notag  \\
	\!\! &=& \!\! \frac{p_{0}^{2}}{2m}+\frac{m\omega ^{2}}{2}\left[ (1+\tau
	p_{0}^{2})^{2}x_{0}^{2}+2i\hbar \tau p_{0}(1+\tau p_{0}^{2})x_{0}\right] ,
	\notag
\end{eqnarray} 
which is obviously non-Hermitian. However, one needs to be careful as the commutation relations (\ref{12}) are also recovered \cite{HrepvsnHrep} from the less obvious Hermitian representations 
\begin{equation}
   X=X^\dagger=x_0,  \quad \text{and} \quad P=P^\dagger=\frac{1}{\sqrt{\tau}} \tan(\sqrt{\tau} p_0) .
\end{equation}
In this case the Hamiltonian (\ref{3}) remains Hermitian 
\begin{equation}
		H_{ho}(X,P) = \frac{1}{2 m\tau} \tan^2(\sqrt{\tau} p_0) + +\frac{m\omega ^{2}}{2}x_0^{2} . \label{HermH}
\end{equation} 
When suitable changing the inner product for non-Hermitian Hamiltonian (\ref{3}), as we shall discuss in general in section \ref{framwork}, it produces the same physics as the Hermitian Hamiltonian (\ref{HermH}) with the standard inner product, see \cite{HrepvsnHrep} for details. Two and three dimensional versions of deformed space-time structures leading to generalised Heisenberg uncertainty relations are discussed in \cite{AFLGFGS,DF5,AFLGBB,DFG,sanjibthesis}.
	
	\subsection{Spectral analysis}
	Let us now explain in general why and under which circumstances a spectrum of a non-Hermitian Hamiltonian may still be real. 
	\subsubsection{Real eigenvalues from unbroken ${\cal{PT}}$-symmetry}
	The first reasoning traces back to Wigner \cite{EW} as mentioned above. Given the TDSE (\ref{TISE}) it makes use of the following three properties: One assumes the existence of an operator ${\cal{PT}}$ that $(i)$ commutes with the Hamiltonian H, which could also be any other operator, $(ii)$ shares the eigenstates $\vert \psi \rangle$ with H and $(iii)$ is antilinear    
	\begin{equation}
		(i) \quad \left[ H,\mathcal{PT}\right] =0, \qquad (ii) \quad \mathcal{PT} \vert \psi \rangle  = e^{i \varphi }\vert \psi \rangle, \qquad (iii) \quad \mathcal{PT} \lambda \vert \psi \rangle =  \lambda^\ast \mathcal{PT} \vert \psi \rangle .  \label{assump}
	\end{equation}
The proof based on these three properties is straightforward
	\begin{equation}
	 e^{i \varphi } E \vert \psi \rangle	\, { \overset{(1)}{=} } \, e^{i \varphi } H \vert \psi \rangle \, { \overset{(ii)}{=} } \,   H \mathcal{PT} \vert \psi \rangle \, { \overset{(i)}{=} } \, \mathcal{PT}  H \vert \psi \rangle \, { \overset{(1)}{=} } \, \mathcal{PT}  E \vert \psi \rangle \, { \overset{(iii)}{=} } \, E^\ast \mathcal{PT}   \vert \psi \rangle \, { \overset{(ii)}{=} } \, e^{i \varphi } E^\ast \vert \psi \rangle \Rightarrow E \in \mathbb{R}.
	\end{equation}
From $(ii)$ and $(iii)$ follows also immediately that $\mathcal{PT}$ is an involution, i.e., $\mathcal{PT}^2 = \mathbb{I}$. We stress that in (\ref{assump}) the operator $H$ is not explicitly assumed to be the Hamiltonian, but in principle could be any operator associated to an eigenvalue equation and most importantly $H$ does not have to be Hermitian as this property is never used in the above argument. Moreover we note that the operator $\mathcal{PT}$ has also not been specified. As an example, it could be taken to be a simultaneous parity $\mathcal{P}$ and time-reversal symmetry $\mathcal{T}$ acting in 1+1 dimensions on the coordinate $x$ and momentum $p$ as $ \mathcal{P}: x\rightarrow -x, p\rightarrow -p ; \; \mathcal{T}: x\rightarrow x, p\rightarrow -p , i\rightarrow -i $ so that $\mathcal{PT}:  x\rightarrow -x, p\rightarrow p, i\rightarrow -i$. Crucially one notes that $\mathcal{PT}$ is antilinear, which one can make plausible for instance by demanding that the standard commutation relation $[x,p]=i \hbar$ is preserved under a $\mathcal{PT}$-transformation. It is trivial to see that $H_{BB}$ respects property $(i)$, but $(ii)$ is less obvious unless we have calculated the explicit expressions for the eigenstates $\vert \psi \rangle$. In fact we can see from the region $\varepsilon < 0$ in figure \ref{figBB} that this might not always be the case.

Let us now assume that property $(ii)$ no longer holds, but instead we have $(ii')$ $\mathcal{PT} \vert \psi_1 \rangle  = \vert \psi_2 \rangle$ with their associated eigenvalue equations $H \vert \psi_1 \rangle = E_1 \vert \psi_1 \rangle$, $H \vert \psi_2 \rangle = E_2 \vert \psi_2 \rangle$. It is now easy to see that in this case the eigenvalues are complex conjugate to each other
\begin{equation}
	E_1 \vert \psi_1 \rangle = H \vert \psi_1 \rangle = H \mathcal{PT} \vert \psi_2 \rangle =  \mathcal{PT} H \vert \psi_2 \rangle =  \mathcal{PT} E_2 \vert \psi_2 \rangle =  E_2^\ast \mathcal{PT} \vert \psi_2 \rangle
	=  E_2^\ast \vert \psi_1 \rangle \Rightarrow E_1 =E_2^\ast.
\end{equation} 
Based of these different possibilities one distinguishes three separate regimes: a) The regime with real eigenvalues where $(i)$, $(ii)$ and $(iii)$ hold one refers to as the {\em $\mathcal{PT}$-symmetric} regime, b) The regime with complex conjugate pairs of eigenvalues where $(i)$, $(ii')$ and $(iii)$ hold one refers to as the {\em spontaneously broken} $\mathcal{PT}$ regime and c) The regime with complex, unrelated, eigenvalues in which also property $(i)$ does not hold as the {\em $\mathcal{PT}$-broken} regime. We summarize these possibilities for a typical two level system in figure \ref{figBB} panel (b), where the energies are plotted as functions of a continuous model parameter $\lambda$ that appears in the Hamiltonian. We observe that unlike as in Hermitian systems, where level crossing is famously always avoided \cite{vNW}, in a non-Hermitian system two energy levels may cross each other by continuing into the complex plane. The crossing points where this occurs are referred to as {\em exceptional points} \cite{Kato,berry2004physics,miri2019exceptional}.   
	
		\subsubsection{Pseudo/Quasi-Hermiticity}
		The next possibility to explain the reality of the spectrum is making use of the fact similarity that transformations do not alter eigenvalue spectra and that Hermitian Hamiltonians are guaranteed to have real eigenvalues as we have seen in (\ref{realEV}). Thus, let us now assume that a non-Hermitian Hamiltonian, $H \neq H^\dagger$, is related to a Hermitian Hamiltonian, $h = h^\dagger$, by means of the adjoint action of a map $\eta$ as
		\begin{equation}
			h = \eta H \eta^{-1} .   \label{Dysonequ}
		\end{equation}	
	The map $\eta$ is often referred to as {\em  Dyson map} and equation (\ref{Dysonequ}) as the {\em  Dyson equation}, because it appears first in a paper by Dyson \cite{Dyson} in 1956. Starting from the time-independent Sch\"odinger equation for the Hermitian Hamiltonian $h$ with real eigenvalue $E$, it is easily seen that $H$ has the same real eigenvalue as long as the Dyson map is well-defined 
		\begin{equation}
		E \vert \phi \rangle  =h \vert \phi \rangle =\eta H \eta^{-1} \vert \phi \rangle \Rightarrow H \eta^{-1} \vert \phi \rangle = E \eta^{-1} \vert \phi \rangle \Leftrightarrow H \vert \psi \rangle = E \vert \psi \rangle, \,\,\,\, \text{with} \,\, \vert \psi \rangle := \eta^{-1}  \vert \phi \rangle .  \label{phiphi}
	\end{equation}	
From the discussion in the previous subsection we expect the Dyson map to be only well-defined in the $\mathcal{PT}$-symmetric regime and to break down when the exceptional point is crossed into the spontaneously broken regime. Another important object that can be calculated directly from the Dyson map is what will turn out below to be the Hermitian {\em metric operator} $\rho=\rho^\dagger$
\begin{equation}
	h=\eta H\eta ^{-1}=(\eta ^{-1})^{\dagger }H^{\dagger }\eta
		^{\dagger }=h^{\dagger }~~\Leftrightarrow ~~ \rho H = H^{\dagger }\rho  \qquad \rho
	:=\eta ^{\dagger }\eta  . \label{1} 
\end{equation} 
When searching the literature there is a slight distinction in terminology depending on the properties of $\rho$. When $\rho$ is positive but not necessarily invertible some authors, especially in the early mathematical literature, refer to the system as being {\em quasi-Hermitian} \cite{Dieu,Urubu}, whereas when $\rho$ is not positive but invertible one says the system is {\em pseudo-Hermitian} \cite{pseudo1,pseudo2,mostafazadeh2002pseudo,Mostafazadeh:2001nr}. In most applications one assumes $\rho$ to be pseudo/quasi-Hermitian that is positive and invertible. We will refer here to the intertwining relation in (\ref{1}) as the {\em quasi-Hermiticity equation}.  

		\subsubsection{Supersymmetry (Darboux transformations)} \label{Darbouxsec}
		Yet another possibility to obtain two isospectral Hamiltonians for which one of them is Hermitian and the other non-Hermitian is making use of a general scheme corresponding to a quantum mechanical analogue of supersymmetry \cite{Witten:1981nf,Cooper:1982dm,Witten,Cooper,bagchi2000super}, also referred to as Darboux-transformations \cite{darboux,crum} in the mathematical literature. 
		We recall the general scheme first and then specify the condition for the isospectral pair to be of the desired type with one being Hermitian and the other non-Hermitian. The starting point is a Hamiltonian containing the analogue of a bosonic and fermionic sector
		\begin{equation}
		\mathcal{H}=H_{1}\oplus H_{2}= L_+ L_- \oplus L_- L_+  ,   \label{Hbf}
		\end{equation} 
		with the two  sub-Hamiltonians factorising into intertwining operators $L_{\pm}$ as specified in (\ref{Hbf}). Evidently we then have the intertwining relations $L_- H_1 = H_2 L_- $ and $L_+ H_2 = H_1 L_+ $. One can then define two operators that play the role of supercharges
		\begin{equation}
			Q:= \left(
			 \begin{array}{cc}
			 	0 & 0 \\
			 	L_- & 0 \\
			 \end{array}
			 \right) \qquad \text{and} \qquad \tilde{Q} := \left(
			 \begin{array}{cc}
			 	0 & L_+ \\
			 	0 & 0 \\
			 \end{array}
			 \right) ,    \label{Scharges}
		\end{equation} 
	obeying the superalgebra $sl(1 \vert 1)$ involving commutation and anti-commutation relations
		\begin{equation}
		[\mathcal{H},Q]=[\mathcal{H},\tilde{Q}]=0, \qquad \{Q, \tilde{Q} \} = \mathcal{H}, \qquad \{Q, Q \} = \{\tilde{Q}, \tilde{Q} \} =0 .
		\end{equation}
		Since the two supercharges commute with the Hamiltonian, Schur's Lemma predicts some degeneracy in the spectra of $H_1$ and $H_2$. Let us see this in more detail and represent the intertwining operators as first order differential operators in the form
		\begin{equation}
			L_{\pm }:=\mp \frac{d}{dx}+W(x), \label{LW}
		\end{equation}
	  where $W(x)$ is referred to as the {\em superpotential}. Assuming further that $H_1$ and $H_2$ have discrete spectra with eigenvalues $E_{n}^{(1)}$, $E_{n}^{(2)}$, respectively, we have
	  \begin{eqnarray}
	  	H_{1}\Phi _{n}^{(1)}\!\!&=&\!\!\left( -\frac{d^{2}}{dx^{2}}+V_{1}\right) \Phi
	  	_{n}^{(1)}=L_{+}L_{-}\Phi _{n}^{(1)}=E_{n}^{(1)}\Phi _{n}^{(1)}, \label{H1}\\
	  		H_{2}\Phi _{m}^{(2)}\!\!&=&\!\!\left( -\frac{d^{2}}{dx^{2}}+V_{2}\right) \Phi
	  	_{m}^{(2)}=L_{-}L_{+}\Phi _{m}^{(2)}=E_{m}^{(2)}\Phi _{m}^{(2)}, \label{H2}
	  \end{eqnarray}
	  with $W(x)$ related to the two potentials as $V_{1}=W^{2}-W^{\prime }$ and $V_{2}=W^{2}+W^{\prime }$.
	   Using next the intertwining relations we observe 
		\begin{eqnarray}
			H_{2}\left( L_{-}\Phi _{n}^{(1)}\right)\!\!&=&\!\!L_{-}H_1\Phi
			_{n}^{(1)}=E_{n}^{(1)}\left( L_{-}\Phi _{n}^{(1)}\right) , \\
			H_{1}\left( L_{+}\Phi _{m}^{(2)}\right)  \!\!&=&\!\!L_{+}H_2\Phi
			_{m}^{(2)}=E_{m}^{(2)}\left( L_{+}\Phi _{m}^{(2)}\right) .
		\end{eqnarray}%
		When comparing these equations with (\ref{H1}), (\ref{H2}) we conclude%
		\begin{equation}
			\Phi _{n+1}^{(1)}=N_{n+1}^{(1)}L_{+}\Phi _{n}^{(2)},~~\ ~\Phi
			_{n}^{(2)}=N_{n}^{(2)}L_{-}\Phi _{n}^{(1)},~~\
			~E_{n}^{(2)}=E_{n+1}^{(1)},~~\ ~E_{0}^{(1)}=0,~~~n\in \mathbb{N}_{0},
			\label{Super}
		\end{equation}
	  with $N_{n}^{(1)}$, $N_{n}^{(2)}$ being some normalisation constants. We notice further that when $L_{-}^{\dagger
	  }=L_{+}$ one may take the normalisation constants simply to $N_{n+1}^{(1)}=1/\sqrt{%
	  	E_{n}^{(2)}}$, $N_{n}^{(2)}=1/\sqrt{E_{n+1}^{(1)}}$. However, as we want to take at least one of our potentials $V_1$ or $V_2$ to be complex, also the superpotential $W$ will be complex and we therefore do not make this assumption. Here the key point is that, apart from the lowest level at $n=0$, the two Hamiltonians $H_{1}$ and $H_{2}$ are
		isospectral. Separating now the superpotential in its real and imaginary part $W=W_r+iW_i$, it is now easy to verify that when	
		\begin{equation}
		          W_r = \pm \frac{1}{2} \partial_x \ln W_i
		\end{equation}
	    holds, we can design a set up in which one of the potentials is real $V_{1/2} \in \mathbb{R}$ and the other is genuinely complex $V_{2/1} \in \mathbb{CR}$.  
		
	\subsection{Quantum mechanical framework} \label{framwork}
	Having established under which circumstances the spectra for non-Hermitian Hamiltonians might be real or partially real, we still need to set up a proper quantum mechanical framework to make sense of these kind of theories.	
		\subsubsection{$H$ is Hermitian with respect to a new metric}
		The key ingredient in the formulation of a consistent ${\cal PT}$-symmetric quasi/pseudo Hermitian quantum mechanical framework is to defined a new inner product for the non-Hermitian theory by means of a new metric. For two eigenstates $\vert \psi \rangle$ and $\vert \tilde{\psi} \rangle$ of the non-Hermitian Hamiltonian $H$ we define the {\em $\rho$-inner product}
		\begin{equation}
		     \left\langle  \psi \right\vert \tilde{\psi} \rangle_\rho :=  \left\langle  \psi \right\vert  \rho \tilde{\psi} \rangle , \label{rhoinner}
		\end{equation}
		where the {\em metric operator} $\rho$ is factorised into the Dyson map as specified in (\ref{1}). The $\rho$-inner product is designed in such a way that $H$ becomes Hermitian (self-adjoint, symmetric) with respect to it, meaning that within the new inner product its action on a ket state equals its action on a bra state. This is seen as follows
		\begin{eqnarray}
			\left\langle  \psi \right\vert H \tilde{\psi} \rangle_\rho \!\!\! &{ \overset{(\ref{1})}{=} }& \!\!\!  \left\langle  \psi \right\vert  \eta^\dagger \eta H \tilde{\psi} \rangle \, { \overset{(\ref{phiphi})}{=} } \, \left\langle \eta^{-1} \phi \right\vert  \eta^\dagger \eta H  \eta^{-1} \tilde{\phi} \rangle
			\, { \overset{(\ref{Dysonequ})}{=} } \,
			\left\langle  \phi \right\vert  h \tilde{\phi} \rangle
			\, { \overset{(\ref{1})}{=} } \,
			\left\langle h \phi \right\vert  \tilde{\phi} \rangle
			\, { \overset{(\ref{Dysonequ})}{=} } \,
				\left\langle \eta H \eta^{-1} \phi \right\vert  \tilde{\phi} \rangle  \quad   \label{23}\\
				 &{ \overset{(\ref{1})}{=} }& \!\!\! \left\langle H \psi \right\vert \tilde{\psi} \rangle_\rho \notag .
		\end{eqnarray}
Here we ignore all domain issues of the operators involved, which of course need to be made precise for a rigorous treatment. We will also keep $x$ and $t$ real and will not discuss how to continue the domains to the complex plane so that one requires to introduce Stokes wedges to ensure a well-defined asymptotic behaviour. 
		\subsubsection{Orthogonality}
		Assuming $H$ to possess a discrete spectrum and given relation (\ref{23}), we may now establish the orthogonality of the corresponding eigenstates when using the $\rho$-inner product. We have
		\begin{equation}
		\left. \begin{array}{c}
			\left\langle \psi _{n}\!\right. \left\vert H\psi _{m}\right\rangle_\rho
			=E _{m} \left\langle \psi _{n}\!\right. \left\vert \psi
			_{m}\right\rangle_\rho  \\
			\left\langle H\psi _{n}\right. \! \vert   \! \left. \psi _{m}\right\rangle_\rho
			=E _{n}^{\ast }\left\langle \psi _{n} \right\vert
			\! \left. \psi_{m}  \right\rangle_\rho 
		\end{array}  \right\}\Rightarrow 0 =(E_{m}-E _{n}^{\ast })\left\langle
		\psi _{n}\!\right. \left\vert \psi _{m}  \right\rangle_\rho ,
		\end{equation}
	so that $\left\langle
	\psi _{n}\!\right. \left\vert \psi _{m}  \right\rangle_\rho=0$ for $n \neq m$. Therefore when normalising the states we obtain the orthonormality relation 
		\begin{equation}
		\left\langle
		\psi _{n}\!\right. \left\vert \psi _{m}  \right\rangle_\rho= \delta_{n,m} .
	\end{equation}
	Note that this re-definition of the inner product is essential as in general the standard inner product is indefinite,  $	\left\langle
	\psi _{n}\!\right.  \left\vert \psi _{m}  \right\rangle_\rho \neq \delta_{n,m}$, and $H$ would be a non self-adjoint operator.
		\subsubsection{How to define observables?}
	To make physical sense of the theory we also need to specify how to identify operators that correspond to physical quantities. We recall that according to the standard axioms of quantum mechanics \cite{von2018mathematical}, observables $o$ are self-adjoint operators acting in a Hilbert space
	\begin{equation}
	\langle \phi  \!\vert o \tilde{\phi}\rangle
	=\langle o \psi   \!\vert \tilde{\phi} \rangle . \label{obsh}
	\end{equation}
In analogy, we define observables $\mathcal{O}$ in the non-Hermitian theory as self-adjoint operators acting in the Hilbert space equipped with the $\rho$-inner product
\begin{equation}
	\left\langle  \psi \right\vert \mathcal{O} \tilde{\psi} \rangle_\rho = \left\langle \mathcal{O} \psi \right\vert \tilde{\psi} \rangle_\rho . \label{obsH}
\end{equation}
Comparing (\ref{obsh}) and (\ref{obsH}) we immediately deduce that the observables $\mathcal{O}$ in the non-Hermitian system must be pseudo/quasi-Hermitian with regard to the observables $o$ in the Hermitian system
	\begin{equation}
\mathcal{O}=\eta^{-1}o\eta\qquad\Leftrightarrow\qquad\mathcal{O}^{\dagger}=
\rho \mathcal{O} \rho^{-1}, \label{obsP}
\end{equation}
just in the same way as $H$ is related to $h$ as stated in (\ref{Dysonequ}).

A little warning is order here. The relation in (\ref{obsP}) imply for instance that the operators $x$ and $p$ occurring in the Hamiltonian $H_{BB}$ in (\ref{HBB}) are in general not observables and can not be interpreted directly as coordinates or momenta. Instead they are mere auxiliary operators and only when the Dyson map and has been constructed can one define the coordinate and momentum operators in the non-Hermitian system as $X:= \eta^{-1} x \eta$ and  $P:= \eta^{-1} p \eta$, respectively. Thus, a direct analysis of the non-Hermitian system leads to a variety of apparent inconsistencies that are not really present when interpreted correctly. Unfortunately many such apparent puzzles can be found in the literature. This confusion persists also in the quantum field theoretical setting where such type of pseudo-problems emerge when one directly analyses the non-Hermitian theory with an incorrect conceptual viewpoint.   		
		
		\subsubsection{General technique, construction of metric and Dyson operators} \label{general}
		As should be clear from the above, the central quantities that one has to determine in the non-Hermitian theory are the Dyson map and the metric operator. In general, this is a difficult task and even for the $H_{BB}$ in (\ref{HBB}) with $\varepsilon = 1$ only perturbative results are known \cite{bender2004scalar,BBJcubic,Mostafazadeh:2004qh,CA,siegl2012metric}. However, many exact solutions for different types of systems with an infinite dimensional Hilbert space have been found \cite{JM,ACIso,Mostsyme,PEGAAF,MGH,PEGAAF2,paulothesis,moniquethesis}. The usual starting point is a given non-Hermitian Hamiltonian $H$. Thus one may commence the construction either by first solving the pseudo/quasi Hermiticity relation (\ref{Dysonequ}) for $\eta$ and subsequently calculate directly the metric operator $\rho$ from (\ref{phiphi}). Alternatively one may also solve the quasi-Hermiticity relation in (\ref{phiphi}) for the metric operator $\rho$ first. At times this is easier, but the subsequent second step of taking the square root in order to obtain the Dyson map $\eta=\sqrt{\rho}$ can be awkward as already stated by Dyson \cite{Dyson}. 
		
		In either case one requires a good Ansatz for the Dyson map or the metric in order to find exact expressions. This is straightforward for a situation in which the Hamiltonian can be expressed in terms of the generators, say $K_i$ of a Lie algebra, as one can simply take $\eta = \exp(\sum_i a_i K_i)$ or $\rho = \exp(\sum_i b_i K_i)$ where the sum extends over the entire rank of the algebra and $a_i$, $b_i$ are constants that need to be determined. Since the adjoint action of such elements on $H$ will produce only expressions expanded in the algebra one obtains a well defined set of equations that in principle can be solved. However, sometimes the algebra is very large, or even infinite dimensional, so that such entirely generic Ansatz might lead to a very complex set of equations or even worse the Hamiltonian can not be expressed in terms of generators of a closed algebra at all. 
		
		In these cases one can resort to perturbation theory \cite{bender2004scalar,BBJcubic,Mostafazadeh:2004qh,CA,siegl2012metric}. To start with we split the non-Hermitian Hamiltonian into its real and imaginary part
		\begin{equation}
			H=h_{0}+i\epsilon h_{1},\qquad \ \ \text{with}~h_{0}^{\dagger
			}=h_{0},h_{1}^{\dagger }=h_{1},  \epsilon \ll 1 .
		\end{equation} 
		For simplicity we assume here that $\eta$ is Hermitian so that $\rho	= \eta ^{\dagger }\eta =\eta ^{2}$ and we take $\rho=e^{q}$ for unknown operators $q$. Using the Baker-Campbell-Hausdorff (BCH) formula\footnote{The adjoint action of $e^A$ on $B$ for two noncommutative operators $A$ and $B$ is given by the BCH formula$$
			e^{A}Be^{-A}=B+[A,B]+\frac{1}{2!}[A,[A,B]]+\frac{1}{3!}[A,[A,[A,B]]]+...$$} we can evaluate the right hand side of the quasi-Hermiticity relation in (\ref{phiphi}) to 
		\begin{equation}
			H^{\dagger }=\eta ^{2}H\eta ^{-2}=H+[q,H]+\frac{1}{2!}[q,[q,H]]+\frac{1}{3!}%
			[q,[q,[q,H]]]+...  \label{eq:sim}
		\end{equation} 
		which can be written as
		\begin{equation}
				i[q,h_{0}]+\frac{i}{2}[q,[q,h_{0}]]+\frac{i}{3!}[q,[q,[q,h_{0}]]]+...=%
				\epsilon \left( 2h_{1}+[q,h_{1}]+\frac{1}{2}[q,[q,h_{1}]]+...\right) .
		\end{equation}
	   Expanding $q$ further as $q=\sum_{n=1}^{\infty }\epsilon ^{n}\check{q}_{n} $ we identify terms order by order in $\epsilon$. Remarkably this set of constraints can be solved recursively. To the lowest order we read off the equations 
		\begin{eqnarray}
			\epsilon^1 \!\!\!&:& \!\!\!\lbrack h_{0},\check{q}_{1}] =2ih_{1},   \label{order1}\\
			\epsilon^3\!\!\!&:& \!\!\!\lbrack h_{0},\check{q}_{3}] =\frac{i}{6}[\check{q}_{1},[\check{q}%
			_{1},h_{1}]], \label{order3}  \\ 
			\epsilon^5\!\!\!&:& \!\!\!\lbrack h_{0},\check{q}_{5}] =\frac{i}{6}\left( [\check{q}_{1},[\check{q}%
			_{3},h_{1}]]+[\check{q}_{3},[\check{q}_{1},h_{1}]]-\frac{1}{60}[\check{q}%
			_{1},[\check{q}_{1},[\check{q}_{1},[\check{q}_{1},h_{1}]]]]\right) .\label{order5}
		\end{eqnarray} 
	We observe now that at first order in $\epsilon$, (\ref{order1}), with known $h_0$ and $h_1$, the only unknown quantity in this equation is $\check{q}_{1}$. Up to the ambiguity of quantities that commute with $h_0$ we can therefore determine $\check{q}_{1}$ form this equation. Similarly at the next nontrivial order $\epsilon^3$, (\ref{order3}), the only unknown quantity that enters the equation is $\check{q}_{3}$, at order $\epsilon^5$, (\ref{order5}) only $\check{q}_{5}$ is not known etc. Thus, these equations may be solved systematically order by order. In many examples the series has been found to terminate so that one has in fact obtained an exact solution and can set $\epsilon=1$.  
	
	\subsubsection{An equivalent approach, biorthonormal basis and the $\mathcal{CPT}$-inner product} \label{bior}
	Since $H$ is non-Hermitian its left and right eigenvectors, say $ \vert \Phi \rangle$ and $\vert \Psi \rangle$, respectively, are not identical with 
	\begin{equation}
	H \vert \Psi_n \rangle = E_n \vert \Psi_n \rangle , \qquad H^\dagger \vert \Phi_n \rangle = E_n \vert \Phi_n \rangle, \qquad n \in \mathbb{N} .  \label{leftright}
	\end{equation}
While in general these eigenvectors are not orthonormal, i.e. $ \langle \Psi_n \vert \Psi_m \rangle \neq \delta_{nm}$, $ \langle \Phi_n \vert \Phi_m \rangle \neq \delta_{nm}$, they do form a {\em biorthonormal basis} \cite{dieudonne1953bio} 
	\begin{equation}
		 \langle \Phi_n \vert \Psi_m \rangle =  \langle \Psi_n \vert \Phi_m \rangle = \delta_{nm}, \qquad \sum_n \vert \Phi_n \rangle \langle \Psi_n \vert = \sum_n \vert \Psi_n \rangle \langle \Phi_n \vert =\mathbb{I} , \label{36}
	\end{equation}  
which has been utilised for instance in the context of dissipative systems, such as in nuclear physics for a long time, see e.g. \cite{moiseyev2011non}. We stress once more that here we are not considering dissipative systems, but instead self-consistent non-Hermitian systems. Next we demonstrate how to convert right eigenstates into left eigenstates by means of the action of the not necessarily positive parity operator
\begin{equation}
	{\cal{P}}  \vert \Psi_n \rangle = s_n  \vert \Phi_n \rangle\qquad s_n = \pm 1 ,  \label{PLR}
\end{equation}
with properties
\begin{equation}
	H^\dagger = {\cal{P}} H {\cal{P}}, \qquad {\cal{P}}^2  =\mathbb{I} . \label{PD} 
\end{equation}
To see how (\ref{PLR}) follows from the properties of biortonormal systems we re-write the second equation in (\ref{leftright}) with (\ref{PD}) as
\begin{equation}
	{\cal{P}} H {\cal{P}} \vert \Phi_n \rangle = E_n \vert \Phi_n \rangle \quad \Rightarrow \quad H {\cal{P}} \vert \Phi_n \rangle = E_n   {\cal{P}} \vert   \Phi_n \rangle . \label{PHP}
\end{equation}
Comparing the last equation in (\ref{PHP}) with the first equation (\ref{leftright}), we conclude that the states ${\cal{P}} \vert   \Phi_n \rangle$ and $\vert \Psi_n \rangle$ must be proportional to each other, so that $ \lambda_n \vert \Psi_n \rangle = {\cal{P}} \vert   \Phi_n \rangle $ with  $ \lambda_n  \in \mathbb{C}$. Moreover, it is easily seen that $\lambda_n$ is in fact real
\begin{equation}
	\lambda_n = \lambda_n \langle \Phi_n \vert \Psi_n \rangle=
	\langle \Phi_n \vert {\cal{P}}  \Phi_n \rangle = 
	\langle \Psi_n \vert \Phi_n \rangle \lambda^\ast = \lambda^\ast .
\end{equation} 
Multiplying next $ \lambda_n \vert \Psi_n \rangle = {\cal{P}} \vert   \Phi_n \rangle $ with its conjugate we derive $\lambda_n^2 =\langle \Psi_n \vert \Psi_n \rangle/\langle \Phi_n \vert \Phi_n \rangle $. Finally we notice that the normalisation condition of the biorthonormal states (\ref{36}) still leave the freedom to scale $\vert \Psi_n \rangle \rightarrow \alpha_n \vert \Psi_n \rangle$, $\vert \Phi_n \rangle \rightarrow \alpha_n^{-1} \vert \Phi_n \rangle$ for arbitrary constants $\alpha_n$. This means that we can always achieve $\lambda_n^2 \rightarrow 1 = s_n^2$, which  in turn  establishes (\ref{PLR}).  

Using the biorthonormal basis one can introduce a new operator  \cite{Bender:2002vv}
\begin{equation}
	{\cal{C}} :=  \sum_n s_n  \vert \Psi_n \rangle \langle \Phi_n \vert, \label{defC}
\end{equation}
where the set $\{s_1, \ldots, s_n\}$ defines the signature. With these quantities one can the identify the metric as the product of the parity ${\cal{P} }$ and ${\cal{C} }$-operator
\begin{equation}
	  \rho= {\cal P C} ,     \label{rPC} 
\end{equation}
and show that the ${\cal CPT}$-inner product introduced in \cite{Bender:2002vv} is in fact identical to the previously discussed $\rho$-inner product
\begin{equation}
	\langle \Psi \vert \tilde{\Psi} \rangle_{{\cal CPT}} := \left( {\cal CPT} \vert \Psi \rangle \right)^\intercal \cdot \vert \tilde{\Psi} \rangle = \langle \Psi \vert {\cal P C} \tilde{\Psi} \rangle =
	\langle \Psi \vert \rho \tilde{\Psi} \rangle
	= \langle \Psi \vert \tilde{\Psi} \rangle_{\rho} .
\end{equation}
For finite dimensional systems the ${\cal{C} }$-operator is often easy to compute so that one can employ the ${\cal CPT}$-inner product upon identifying also the ${\cal{P} }$-operator, but in general one does not have the entire left and right eigenvector set at one's disposal, so that it appears easier to solve the Dyson equation (\ref{Dysonequ}) directly or the quasi-Hermiticity equation (\ref{1}). We also stress here that ${\cal C}$ is not related to charge conjugation in the standard sense, so that also ${\cal CPT}$ does not refer to this well-known symmetry in quantum field theory. In fact, as the standard ${\cal{C} }$-operator maps particles to anti-particles in the same theory, there is no such operator for systems described by the Schr\"odinger equation since it is a single particle description. Note also that while the parity operator ${\cal{P} }$ satisfies the Dyson equation (\ref{Dysonequ}) it is usually not positive definite as we require for a well-defined metric operator $\rho$.

	\subsubsection{An examples with finite dimensional Hilbert space} \label{exfinite}
	 Let us now make our generic discussion more concrete by presenting two explicitly worked out examples. The simplest case consists of a two-level system with a finite dimensional Hilbert space. All Hamiltonians of this type can be expressed in an $SU(2)$-invariant form \cite{Bender:2003gu,PEGAAF2}, of which we select the specific case described by the non-Hermitian Hamiltonian
	\begin{equation}
		H=-\frac{1}{2}\left( \omega \mathbb{I}+\lambda \sigma _{z}+i\kappa \sigma
		_{x}\right)=-\frac{1}{2}\left(
		\begin{array}{cc}
			\omega + \lambda   & i \kappa  \\
			i \kappa  & \omega -\lambda  \\
		\end{array}
		\right), \qquad \omega, \lambda, \kappa \in \mathbb{R},   \label{Hfinite}
	\end{equation}
expressed in terms of the standard Pauli matrices 
	\begin{equation}
\mathbb{I}= \left(
	\begin{array}{cc}
		1 & 0 \\
		0 & 1 \\
	\end{array}
	\right), \qquad \sigma_x = \left(
	\begin{array}{cc}
		0 &1 \\
		1 & 0 \\
	\end{array}
	\right) ,   
	\qquad \sigma_y = \left(
	\begin{array}{cc}
		0 & -i \\
		i & 0 \\
	\end{array}
	\right) , \qquad \sigma_z = \left(
	\begin{array}{cc}
		1 & 0 \\
		0 & -1 \\
	\end{array}
	\right) .\label{Pauli}
\end{equation} 
It is a simple exercise to solve the eigenvalue equation $H \Psi_{\pm } = E_{\pm } \Psi_{\pm } $ for the eigenvalues $E_{\pm }$ and eigenstates $\Psi _{\pm }$ to
\begin{equation}
	E_{\pm }=-\frac{1}{2}\omega \pm \frac{1}{2}\sqrt{\lambda ^{2}-\kappa ^{2}}%
	,\qquad \text{and} \qquad \vert \Psi _{\pm } \rangle= \frac{1}{N_{\pm}}  \left( 
	\begin{array}{c}
		i(-\lambda \pm \sqrt{\lambda ^{2}-\kappa ^{2}}) \\ 
		\kappa%
	\end{array}%
	\right) ,  \label{eigenv}
\end{equation} 
respectively. The normalisation constants  $N_{\pm}= \sqrt{2} \sqrt{\lambda  \sqrt{\lambda ^2-\kappa ^2}\pm\kappa ^2\mp\lambda ^2}$ are real for real energy eigenvalues. By direct inspection we identify the exceptional point at $\lambda = \kappa$, as evidently the spectrum of this Hamiltonian is real for $ \left\vert \lambda \right\vert >\left\vert \kappa \right\vert$ and when $\left\vert \lambda \right\vert < \left\vert
\kappa \right\vert$ the eigenvalues form a complex conjugate pair. The normalisation constants need to be adjusted at the exceptional point when the two eigenvectors coalesce, so that one does not simply encounter a standard degeneracy. The Jordan normal form for the matrix $H$ becomes non-diagonal in this case, so that the exceptional point needs to be treated separately. The $\mathcal{PT}$-symmetry for this system is also easily identified as
\begin{equation}
	\mathcal{PT}:= \sigma _{z} \tau,  \label{PTPT}
\end{equation}  
where $\tau$ denotes here the complex conjugation. We notice that all three properties in (\ref{assump}) that are required to guarantee the reality of the spectrum hold for this map with $\varphi = \pi$ in the $\mathcal{PT}$-symmetric regime  $ \left\vert \lambda \right\vert >\left\vert \kappa \right\vert$. In the spontaneously broken regime property (ii) no longer holds. It is also straightforward to find the Dyson map. Defining the matrix $\eta= \{\psi_+,\psi_-\}^\intercal$ with the eigenvectors of $H$ as column vectors, we verify the Dyson equation (\ref{Dysonequ}) and the quasi-Hermiticity relation (\ref{1}) as
\begin{equation}
	h= \eta H \eta^{-1} = \left( \begin{array}{cc}
		E_+ & 0 \\
		0 & E_- \\
	\end{array} \right), \quad   \rho H \rho^{-1} = H^\dagger \quad  \text{with} \;\; \rho =\eta^\dagger \eta= \frac{1}{\sqrt{\lambda^2 -\kappa^2}}\left(
\begin{array}{cc}
\lambda  & i \kappa  \\
-i \kappa  & \lambda  \\
\end{array}
\right) .  \label{rhoex}
\end{equation}  
The Dyson map as well as the metric break down at the exceptional point $\lambda = \kappa$. We also find that $\det{\rho}=1$ and both eigenvalues of the metric $\left\{{\sqrt{\lambda ^2-\kappa ^2}}/{(\kappa +\lambda) },{(\kappa +\lambda) }/{\sqrt{\lambda ^2-\kappa ^2}}\right\}$ are positive in the $\mathcal{PT}$-symmetric regime.

For completeness we also demonstrate here that we obtain the same metric $\rho$ by utilising the biorthonormal basis and the $\mathcal{C}$-operator as outlined in section \ref{bior}. First of all we recognise the $\mathcal{P}$-operator by identifying the factors in equation (\ref{PTPT}) as $\mathcal{P} = \sigma_z$ and convince ourselves that this operator does indeed satisfy the equations in (\ref{PD}). In this construction we also require the left eigenvectors of $H$, i.e. the eigenvectors of $H^\dagger$  
\begin{equation}
	\vert \Phi _{\pm } \rangle= \frac{1}{N_{\pm}}  \left( 
	\begin{array}{c}
		i( \pm \lambda - \sqrt{\lambda ^{2}-\kappa ^{2}}) \\ 
		\kappa%
	\end{array}%
	\right) .
\end{equation}
Using equation (\ref{PLR}) we identify the signature as $\{s_+,s_-\}= \{-1,1\}$, so that we can calculate the $\mathcal{C}$-operator from its defining relation (\ref{defC}), obtaining
\begin{equation}
{\cal C} =\frac{1}{\sqrt{\lambda^2 -\kappa^2}}\left(
	\begin{array}{cc}
		\lambda  & i \kappa  \\
		i \kappa  & -\lambda  \\
	\end{array}
	\right).
\end{equation}
Thus using (\ref{rPC}) we derive the same metric operator as in (\ref{rhoex}) from the product $\mathcal{PC}=\rho$. We have also seen that both ${\cal P}$ and $\rho$ satisfy the quasi-Hermiticity relation, but are distinct by $\rho$ being positive definite and ${\cal P}$ having positive and negative eigenvalues.

One of the remarkable features we will encounter below is that when introducing a time-dependence via $\lambda \rightarrow \lambda(t)$, $\kappa \rightarrow \kappa (t)$ the time-dependent energies of this system become real and the inner product remain well defined in the broken  $\mathcal{PT}$-regime when $\left\vert \lambda(t) \right\vert < \left\vert \kappa(t) \right\vert$.   
	
	The treatment of time-dependent ${\cal PT}$-symmetric spin chains with larger finite dimensional Hilbert spaces can be found in \cite{CKW,chainOla,Andrei}.  

	\subsubsection{An examples with an infinite dimensional Hilbert space} \label{infnotime}
	Next we consider a non-Hermitian Hamiltonian with an infinite dimensional Hilbert space, a $\mathcal{PT}$-symmetrically coupled harmonic oscillator
	\begin{equation}
		H_{K}=aK_{1}+bK_{2}+i\lambda K_{3},  \qquad a,b,\lambda \in \mathbb{R} \label{HK}
	\end{equation}
	where the operators
	\begin{equation}
		K_{1}=\frac{1}{2}\left( p_{x}^{2}+x^{2}\right),~~K_{2}=\frac{1}{2}\left(
		p_{y}^{2}+y^{2}\right),~~K_{3}=\frac{1}{2}\left( xy+p_{x}p_{y}\right),~~ 
		K_{4}=\frac{1}{2}\left( xp_{y}-yp_{x}\right) 
	\end{equation} 
are defined in terms of standard momenta $p_x, p_y$ and coordinates $x, y$ with non-vanishing canonical commutation relations $[x,p_x]=[y,p_y]=1$. The operators $K_i$, $i=1,2,3,4$ satisfy the closed Lie algebra
	\begin{equation}
	\begin{array}{lll}
		\left[ K_{1},K_{2}\right] =0,~ & \left[ K_{1},K_{3}\right] =iK_{4}, & \left[
		K_{1},K_{4}\right] =-iK_{3}, \\ 
		\left[ K_{2},K_{3}\right] =-iK_{4},~~ & \left[ K_{2},K_{4}\right] =iK_{3},~~
		& \left[ K_{3},K_{4}\right] =\frac{i}{2}(K_{1}-K_{2}).
	\end{array} \label{alg}
\end{equation}
Inspecting the Hamiltonian we identify two antilinear $	\mathcal{PT}$-symmetries
\begin{equation}
	\mathcal{PT}_{\pm }:x\rightarrow \pm x, 
	y\rightarrow \mp y, p_{x}\rightarrow \mp p_{x}, p_{y}\rightarrow \pm
p_{y}, i\rightarrow -i,
\end{equation} 
by demanding $\left[ \mathcal{PT}_{\pm },H_{K}\right] =0$. Strictly speaking we might view these symmetries only as partial $\mathcal{PT}$-symmetries, but it has become quite common to refer to any type of antilinear symmetry as ``$\mathcal{PT}$'' by keeping in mind that this does not necessary mean a total reflection in space and time. Potentially it might not even involve any reflections at all, but simply permutations.

In order to determine the Dyson map $\eta$ from equation (\ref{Dysonequ}) we need to make a suitable Ansatz or use perturbation theory as outlined in section \ref{general}. Since the Hamiltonian is entirely expressed in terms of generators of a closed algebra, we may assume $\eta= \exp( \sum_{i=1}^4 c_i K_i)$ with constants $c_i$ to be determined. Acting adjointly on $H$ with this operator and using the BCH formula we demand the result of this calculation to be Hermitian. Since each term with $K_i$ is linearly independent, the requirement that the non-Hermitian terms must vanish leads to a well-defined system of equations that may be solved, provided such a solution exists. The simplest solution we obtain in this case is $\eta =e^{2 \theta  K_4}$ where $\theta :=\func{arctanh}[\lambda /(b-a)]$ with isospectral Hermitian counterpart
		\begin{equation}
		h_{K}=\frac{1}{2}(a+b)\left( K_{1}+K_{2}\right) +\frac{1%
		}{2}\sqrt{(a-b)^{2}-\lambda ^{2}}\left( K_{1}-K_{2}\right) .
	\end{equation}
Alternatively we could have used perturbation theory with $h_0 = a K_{1}+b K_{2}$,  $h_1 =K_3$ and $\epsilon = \lambda$. The first order equation (\ref{order1}) is then easily solved to $\check{q}_1 =2/(b-a) K_4$, from the third order equation (\ref{order3}) we obtain $\check{q}_3 =2/3(b-a)^3 K_4$, the fifth order equation (\ref{order5}) is solved by $\check{q}_5 =2/5(b-a)^5 K_4$, etc. Computing enough terms one may then extrapolate to all orders in $\lambda$ and try to express $q$ in terms of a standard function. In this case we recognise that the first terms correspond to the Taylor expansion of $2 \func{arctanh}[\lambda /(b-a)]$ in $\lambda$ about $\lambda=0$.  

From the Dyson map and the form of the Hermitian Hamiltonian $h_K$, it is already clear, even without the inspection of the spectrum, that the exceptional point is at $ \vert \lambda \vert = \vert b-a \vert$ and when $ \vert \lambda \vert > \vert b-a \vert$ the $\mathcal{PT}$-symmetry is spontaneously broken. Thus taking now $a=b$ we are always in the broken regime unless $\lambda =0 $ for which $H_K$ is obviously Hermitian. We confirm this with the explicit computation of the eigensystem. In this case we find a discrete spectrum of eigenenergies  
\begin{equation}
	E_{n,m}=E_{m,n}^{\ast }=a(1+n+m)+i\frac{\lambda }{2}(n-m),
\end{equation}
and eigenfunctions
	\begin{equation}
	\psi _{n,m}(x,y) = \frac{e^{-\frac{x^{2}}{2}-\frac{y^{2}}{2}}}{2^{n+m}\sqrt{%
			n!m!\pi }}   \left[ \dsum\limits_{k=0}^{n}\binom{n}{k}H_{k}(x)H_{n-k}(y)\right]  
		 \left[ \dsum\limits_{l=0}^{m}(-1)^{l}\binom{m}{l}H_{l}(y)H_{m-l}(x)\right],
\end{equation}
with $H_n(x)$ denoting the $n$th Hermite polynimial. As expected the energy spectrum consists mainly of pairs of complex conjugate eigenvalues unless $n=m$, in which case we also restore the $\mathcal{PT}$-symmetry by noting that $\mathcal{PT}_{\pm} \psi _{n,m}(x,y) =\psi _{n,m}(x,y) $. It may appear somewhat unmotivated at this point to consider the spontaneously broken regime. However, our main reason to do this here is that we will investigate this model once more in the time-dependent case for which we find that the energy spectrum becomes real for any $a \rightarrow a(t)$, $ \lambda \rightarrow  \lambda(t)$.

\section{${\cal{PT}}$-symmetric quantum mechanics - time-dependent $H(t)$}
As mentioned in the introduction, many concrete Hamiltonian systems can not be described by autonomous Hamiltonians $H$, but require an explicit dependence on time $H(t)$. In this part of the lecture we discuss how such type of systems can be treated consistently when $H(t)$ is non-Hermitian, that is $H(t) \neq H^{\dagger}(t)$. 
	\subsection{Theoretical framework (key equations)} 
	To start with we assume also the existence of an explicitly time-dependent Hermitian Hamiltonian $h(t)=h^{\dagger}(t)$. Each of these Hamiltonians satisfied their own respective TDSE
	\begin{equation}
		h(t)\phi (t)=i\hbar \partial _{t}\phi (t),\qquad \text{and\qquad }H(t)\Psi
		(t)=i\hbar \partial _{t}\Psi (t) . \label{TDSEhH}
	\end{equation}
Similarly as in the time-independent scenario we assume the respective wave functions to be related to each other as 	
\begin{equation}
	\phi (t)=\eta (t)\Psi (t), \label{phipsi}
\end{equation} 
with the difference that the Dyson map is now also time-dependent.	Substituting (\ref{phipsi}) into the RHS of the TDSE for $h(t)$ and subsequently using the TDSE for $H(t)$ we derive the {\em time-dependent Dyson equation} (TDDE)
	\begin{equation}
		h(t)=\eta (t)H(t)\eta ^{-1}(t)+i\hbar \partial _{t}\eta (t)\eta ^{-1}(t) . \label{TDDE}
\end{equation} 
For time-independent Dyson map this equation reduces to the standard Dyson equation (\ref{Dysonequ}). Equation (\ref{TDDE}) reminds on a gauge transformation that relates two different time-dependent Hamiltonians by a gauge transformation, see e.g. (2.8) in \cite{AC1}, but the crucial difference here is that the Dyson map is not a unitary operator so that (\ref{phipsi}) is not a gauge transformation. We therefore say the last term in (\ref{TDDE}) is ``gauge-like''.

Similarly we generalise the quasi-Hermiticity equation (\ref{1}) to the {\em time-dependent quasi-Hermiticity equation} (TDQHE)
\begin{equation}
		H^{\dagger }\rho (t)-\rho (t)H=i\hbar \partial _{t}\rho (t), \label{TDQHE}
\end{equation}
by conjugating (\ref{TDDE}) and using the definition $\rho (t):=\eta^{\dagger }(t)\eta (t)$ for the time-dependent metric. Thus in complete analogy to (\ref{rhoinner}) we interpret 
\begin{equation}
	\left\langle  \psi(t) \right \vert \tilde{\psi}(t) \rangle_\rho :=  \left\langle  \psi(t) \right\vert  \rho(t) \tilde{\psi}(t) \rangle  \label{rhoinnert}
\end{equation}
as the time-dependent $\rho$-inner product between two states $	\left\langle  \psi(t) \right\vert$ and $ \vert \tilde{\psi}(t) \rangle$.
Consequently, as observables $o(t)$ in the Hermitian system are self-adjoint operators acting in a Hilbert space, the observables $\mathcal{O}(t) $ in the non-Hermitian system must be quasi-Hermitian, i.e., they have to satisfy 
\begin{equation}
	o(t)=\eta (t)\mathcal{O}(t)\eta ^{-1}(t)  . \label{oO}
\end{equation} 
For expectation values we therefore have 
	\begin{equation}
	\left\langle \phi (t)\right. \left\vert o(t) \phi (t)\right\rangle
	=\left\langle \Psi (t)\left\vert \rho (t)\mathcal{O}(t) \Psi (t)\right\rangle
	\right.  =  \left\langle \Psi (t)\left\vert \mathcal{O}(t) \Psi (t)\right\rangle
	\right._\rho .  \label{expt}
\end{equation}
Thus once again the central objects are the time-dependent Dyson map $\eta(t)$ and the time-dependent metric $\rho(t)$, which we obtain by solving the TDDE (\ref{TDDE}) or the TDQHE (\ref{TDQHE}), respectively.
	
	\subsection{The nondual nature of the Hamiltonian}
	For Hermitian Hamiltonians, time-independent $h$ as well as time-dependent $h(t)$, one is used to the fact that the Hamiltonian governs the time-evolution of the system and simultaneously is also the operator that corresponds to the observable energy, time-independent $E$ as well as time-dependent $E(t)$. While this dual nature of the Hamiltonian still holds for time-independent non-Hermitian Hamiltonians $H$, it is lost for their time-dependent versions $H(t)$. This somewhat unexpected feature has led to some controversy and debate questioning even the possibility to consistently set up time-dependent non-Hermitian Hamiltonian systems. While some of the confusion originated from linguistic ambiguities, the resolution is to make a clear conceptual distinction between the Hamiltonian $H(t)$, which is the operator that satisfies the TDSE, therefore governing the time evolution of the system, and the energy operator, say $\tilde{H}(t)$, that is quasi-Hermitian with respect to a Hermitian Hamiltonian/energy operator $h(t)$.  
	
	Thus taking the Hamiltonian $H(t)$ in (\ref{TDSEhH}) as starting point, its nonquasi-Hermitian nature is immediately clear from (\ref{TDDE}), so that this operator is not an observable by (\ref{oO}). Instead we can define the observable {\em energy operator} 
	\begin{equation}
		\tilde{H}(t)=\eta ^{-1}(t)h(t)\eta (t)=H(t)+i\hbar \eta ^{-1}(t)\partial
			_{t}\eta (t) . \label{energyop}
	\end{equation}
By the same reasoning as in the time-independent case it follows now directly that $\tilde{H}(t)$ satisfies the quasi-Hermiticity relation $\rho(t)\tilde{H}(t)=\tilde{H}(t)^\dagger.\rho(t)$. 
Of course one could demand $\tilde{H}(t)$ to satisfy its own TDSE, but this would simply lead to a new unrelated Hilbert space and just duplicate the above conundrum. 

Thus with the Hamiltonian $H(t)$ satisfying the TDSE, it governs the time-evolution and must therefore be related to a unitary time-evolution operator $U(t,t^{\prime })$ that transforms a state at time $t^{\prime }$ to a state at time $t$. We recall first the standard properties of this operator $u(t,t^{\prime })$ in the Hermitian case
\begin{equation}
	\phi (t)=u(t,t^{\prime })\phi (t^{\prime }),  \qquad  	u(t,t^{\prime })=T\exp \left[ -\frac{i}{\hbar}\int\nolimits_{t^{\prime }}^{t}dsh(s)\right]  ,
\end{equation} 
where $T$ denotes the time-ordered product. Using the expression for $u(t,t^{\prime })$ and the TDSE for $h(t)$ we easily derive time-evolution operator also satisfied the TDSE
\begin{equation}
	h(t)u(t,t^{\prime })=i\hbar \partial _{t}u(t,t^{\prime }), \qquad 	u(t,t^{\prime })u(t^{\prime },t^{\prime \prime })=u(t,t^{\prime \prime}),  \qquad u(t,t)=\mathbb{I} .  \label{propu}
\end{equation} 
Moreover, as $u(t,t^{\prime })$ is unitary we have
\begin{equation}
	\left\langle u(t,t^{\prime
	})\phi (t^{\prime })\left\vert u(t,t^{\prime })\tilde{\phi}(t^{\prime
	})\right\rangle \right. =\left\langle \phi (t)\left\vert \tilde{\phi}%
	(t)\right\rangle \right. .
\end{equation}  
For the non-Hermitian case we introduce the unitary time-evolution operator $U(t,t^{\prime })$ in an analogous fashion 
\begin{equation}
	\Psi (t)=U(t,t^{\prime })\Psi (t^{\prime }), \qquad U(t,t^{\prime })=T\exp \left[ -\frac{i}{\hbar}\int\nolimits_{t^{\prime }}^{t}dsH(s)\right]  .
\end{equation} 	
Since $H(t)$ satisfies the TDSE we derive the properties corresponding to (\ref{propu}) as 
\begin{equation}
	H(t)U(t,t^{\prime })=i\hbar \partial _{t}U(t,t^{\prime }), \qquad 	U(t,t^{\prime })U(t^{\prime },t^{\prime \prime })=U(t,t^{\prime \prime}),  \qquad U(t,t)=\mathbb{I} . \label{propU}
\end{equation}   
 From (\ref{expt}) it follows now directly that 
\begin{equation}
	\left\langle U(t,t^{\prime })\Psi (t^{\prime })\left\vert U(t,t^{\prime })%
	\tilde{\Psi}(t^{\prime })\right\rangle \right. _{\rho }=\left\langle \Psi
	(t)\left\vert \tilde{\Psi}(t)\right\rangle \right. _{\rho } ,
\end{equation}
and in addition that the time-evolution operator of the Hermitian and non-Hermitian system are related by the adjoint action of the time-dependent Dyson map
	\begin{equation}
	U(t,t^{\prime})=\eta ^{-1}(t)u(t,t^{\prime })\eta (t^{\prime }) . \label{Dysonu}
\end{equation}  
Using these properties we derive directly a generalization of the Du Hamel formula, which is extremely useful in the context of gauge theories and in setting up a perturbation theory, see \cite{reed1972methods,AC1},
	\begin{eqnarray}
	U(t,t^{\prime})\!\!&=& \!\!u(t,t^{\prime}) - \int_{t^{\prime }}^{t} \frac{d}{ds}  \left[  U(t,s)u(s,t^{\prime }) \right]  ds     \\		 
	\!\!&=&\!\!	u(t,t^{\prime})- \frac{i}{\hbar} \int_{t^{\prime }}^{t} U(t,s) \left[ H(s)-h(s) \right] u(s,t^{\prime }) ds  . \label{62}
\end{eqnarray}
When iterating this formula, that is by replacing $U(t,s)$ on the LHS by using (\ref{62}) with $t^{\prime} \rightarrow s$, one obtains a power series for $U(t,t^{\prime})$ in terms of the difference $H(t)-h(t)$. In the Hermitian case the series corresponds to the famous Dyson series, in which each term can be represented by a sum of Feynman diagrams, that expands the time-evolution operator in a series of the interacting potential when we replace $H(t) \rightarrow H_0 +V(t)$ and $h(t) \rightarrow H_0 $.  

The quantum mechanical Green's function of the Hermitian and the non-Hermitian system can be defined in term of the time-evolution operator as
\begin{equation}
	G_h(t,t^{\prime }):= - \frac{i}{\hbar} u(t,t^{\prime }) \theta(t- t^{\prime }), \qquad G_H(t,t^{\prime }):= - \frac{i}{\hbar} U(t,t^{\prime }) \theta(t- t^{\prime }),
\end{equation}
respectively, with $\theta(x)$ denoting the Heaviside step function. Using the properties (\ref{propu}) and (\ref{propU}), we easily verifies that they indeed satisfy the crucial relations
\begin{equation}
  (i \hbar \partial_t - h)	G_h(t,t^{\prime })= \delta (t-t^{\prime }),
  \qquad   (i \hbar \partial_t - H)	G_H(t,t^{\prime })= \delta (t-t^{\prime }).
\end{equation}
The Du Hamel formula that relates the Green's function of the Hermitian and the non-Hermitian system is derived from (\ref{62}) to
\begin{equation}
	G_U(t,t^{\prime}) =	G_u(t,t^{\prime})+  \frac{1}{\hbar} \int_{-\infty}^{\infty} G_U(t,s) \left[ H(s)-h(s) \right] G_u(s,t^{\prime}) ds .
\end{equation}  
	Notice that for the derivations of the Du Hamel formulae we did not make use of (\ref{Dysonu}), so that we can make different choices for $H(t)$ and $h(t)$. For instance, when separating $H(t) =h_0(t) + i h_1(t)$, with $h^\dagger_0(t)=h_0(t)$ $h^\dagger_1(t)=h_1(t)$, we can set up a perturbative expansion in term of the Hermitian term $h_1(t)$ by replacing $h(t) \rightarrow h_1(t)$, provided $ h_1(t) \ll 1$. Of course depending on the relative size of the individual terms we can also make different choices. We will return to a time-dependent version of a pertubative approach in section \ref{pertsec}. 
	
	\subsection{Solution procedures}  
	Next we discuss how to obtain the time-dependent Dyson map $\eta(t)$ and the time-dependent metric $\rho(t)$, by solving the TDDE (\ref{TDDE}) or the TDQHE (\ref{TDQHE}), respectively. Obviously due to the presence of the gauge-like term, this is more involved than in the time-independent scenario where one simply had to find a similarity transformation. Taking the Dyson map for instance to be of the form $\eta(t) = \exp \sum_i \alpha_i(t) q_i$ we do not only have to take care about the commutation relations for the operators $q_i$ but in addition the time-dependent coefficient functions have to satisfy usually a set of coupled first order differential equations.

		\subsubsection{Three scenarios for possible time-dependence} \label{scenarios}
		It is useful to distinguish between three different types of scenarios depending on the explicit time-dependence of the different operators involved.
		\begin{enumerate}
			\item    $\partial_t \eta =0$ ,   $\partial_t H \neq 0$,  $\partial_t h \neq 0$ \\
			For a time-independent Dyson, and hence metric, the gauge-like term in the time-dependent Dyson equation (\ref{TDDE}) vanishes so that this case reduces technically to the time-independent case in which case the time $t$ plays the role as any other parameter of the theory \cite{CA,CArev}.
			\item    $\partial_t \eta \neq 0$ ,   $\partial_t H = 0$,  $\partial_t h \neq 0$\\
			We may also consider a time-independent non-Hermitian Hamiltonian, but seek a solution for the Dyson map that is explicitly time-dependent so that one still needs to solve the full TDDE (\ref{TDDE}). This case was explored in \cite{AndTom2} and opens up the new possibility of the {\em metric picture} as it allows the time-dependence to be included in the metric rather than in the familiar Schr\"{o}dinger picture with time-dependent states, the Heisenberg picture with time-dependent observables or the interaction picture being a mixture between the two later versions.
			\item   $\partial_t \eta \neq 0$ ,   $\partial_t H \neq 0$,  $\partial_t h \neq 0$ \\
		     This case, in which all involved quantities are explicitly time-dependent, will be our main focus below. We will pursue here various possibilities to find the Dyson map. We may either solve directly the full TDDE (\ref{TDDE}) for $\eta(t)$ and subsequently calculate the metric operator directly from $\rho (t):=\eta^{\dagger }(t)\eta (t)$ or solve first the TDQHE (\ref{TDQHE}) and thereafter compute $\eta(t)$ from $\rho (t):=\eta^{\dagger }(t)\eta (t)$. Alternatively, we can employ Lewis-Riesenfeld invariants, as will be discussed in section \ref{LRinv}, which reduces the problem  technically to finding a similarity transformation between two invariants. The challenge in this approach is to find the invariants, which may be attempted in a exact manner with a suitable Ansatz, point transformations or in a semi-exact manner by approximation the solutions to the eigenvalue equations for the invariants. 
		\end{enumerate}	  	
	\subsubsection{Exact time-dependent solutions for a finite dimensional Hilbert space} Let us return to our two-level example treated in section \ref{exfinite}. Keeping the Hamiltonian time-independent we discuss now the scenario (ii) introduced in the previous subsection \ref{scenarios}. Making the most general Ansatz for the time-dependent metric operator
	\begin{equation}
			\rho(t)= \alpha_0(t) \mathbb{I}+ \alpha_1(t) \sigma_x+ \alpha_2(t) \sigma_y+ \alpha_3(t) \sigma_z ,
	\end{equation}
	we try to determine the time-dependent coefficient functions $\alpha_i(t)$ by solving for the TDQHE (\ref{TDQHE}). Upon substituting  $\rho(t)$ into (\ref{TDQHE}) the $\alpha_i$ have to satisfy the following set of coupled first order differential equations 
	\begin{eqnarray}
	 \dot{\alpha} _0\!\!&=&\!\!	\kappa  \alpha _1-\dot{\alpha} _3, \qquad
	 \dot{\alpha} _1 =\alpha _0 \kappa +\left(\alpha _2+i \alpha _1\right) \lambda +i \dot{\alpha} _2, \\
	 \dot{\alpha} _0\!\!&=&\!\!	\kappa  \alpha _1+\dot{\alpha} _3, \qquad
	 \dot{\alpha} _1 =\alpha _0 \kappa +\left(\alpha _2-i \alpha _1\right) \lambda -i \dot{\alpha} _2 .
	\end{eqnarray}
We denote here partial derivatives with respect to $t$ by overdots. With some straightforward manipulations we solve these equations to
\begin{eqnarray}
 \alpha_0(t)\!\!&=&\!\!\frac{\kappa  \left(c_1 e^{t \sqrt{\kappa ^2-\lambda ^2}}-c_2 e^{-t \sqrt{\kappa ^2-\lambda
 			^2}}\right)}{\sqrt{\kappa ^2-\lambda ^2}}+c_3, \qquad
 \alpha_1(t)=c_1 e^{t \sqrt{\kappa ^2-\lambda ^2}}+c_2 e^{-t \sqrt{\kappa ^2-\lambda ^2}}, \qquad \\
	 \alpha_2(t)\!\!&=& \!\!\frac{\lambda  \left(c_2 e^{-t \sqrt{\kappa ^2-\lambda ^2}}-c_1 e^{t \sqrt{\kappa ^2-\lambda
	 			^2}}\right)}{\sqrt{\kappa ^2-\lambda ^2}}-\frac{c_3 \kappa }{\lambda } , \qquad \alpha_3(t)=c_4 ,
\end{eqnarray}
with four integration constants $c_i$, $i=1,\ldots , 4$ as expected. Notice that there is no choice for the $c_i$ that would smoothly connect to the solution (\ref{rhoex}) found for the time-independent case. However, considering $\det[\rho(t)] = c_3^2 (1-\kappa^2/\lambda^2)- 4 c_1 c_2 -c_4$  it is conceivable that for suitable choices of the integration constants we may obtain a positive definite time-dependent metric. 

\begin{figure}[h]
	\noindent	\begin{minipage}[b]{0.50\textwidth}     \!\!\!\! \!\!\!\! \includegraphics[width=\textwidth]{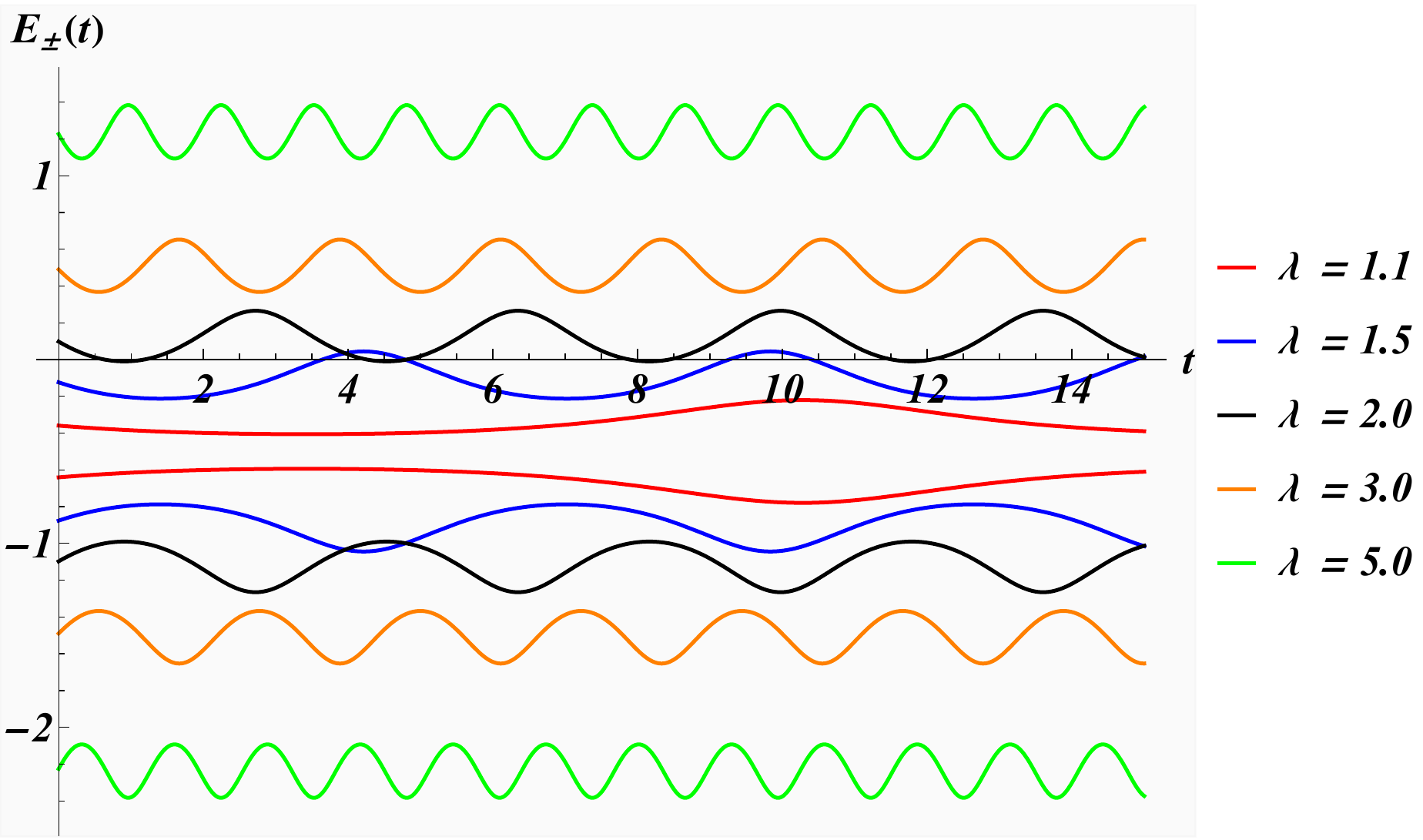}
	\end{minipage}
	\begin{minipage}[b]{0.50\textwidth}      \includegraphics[width=\textwidth]{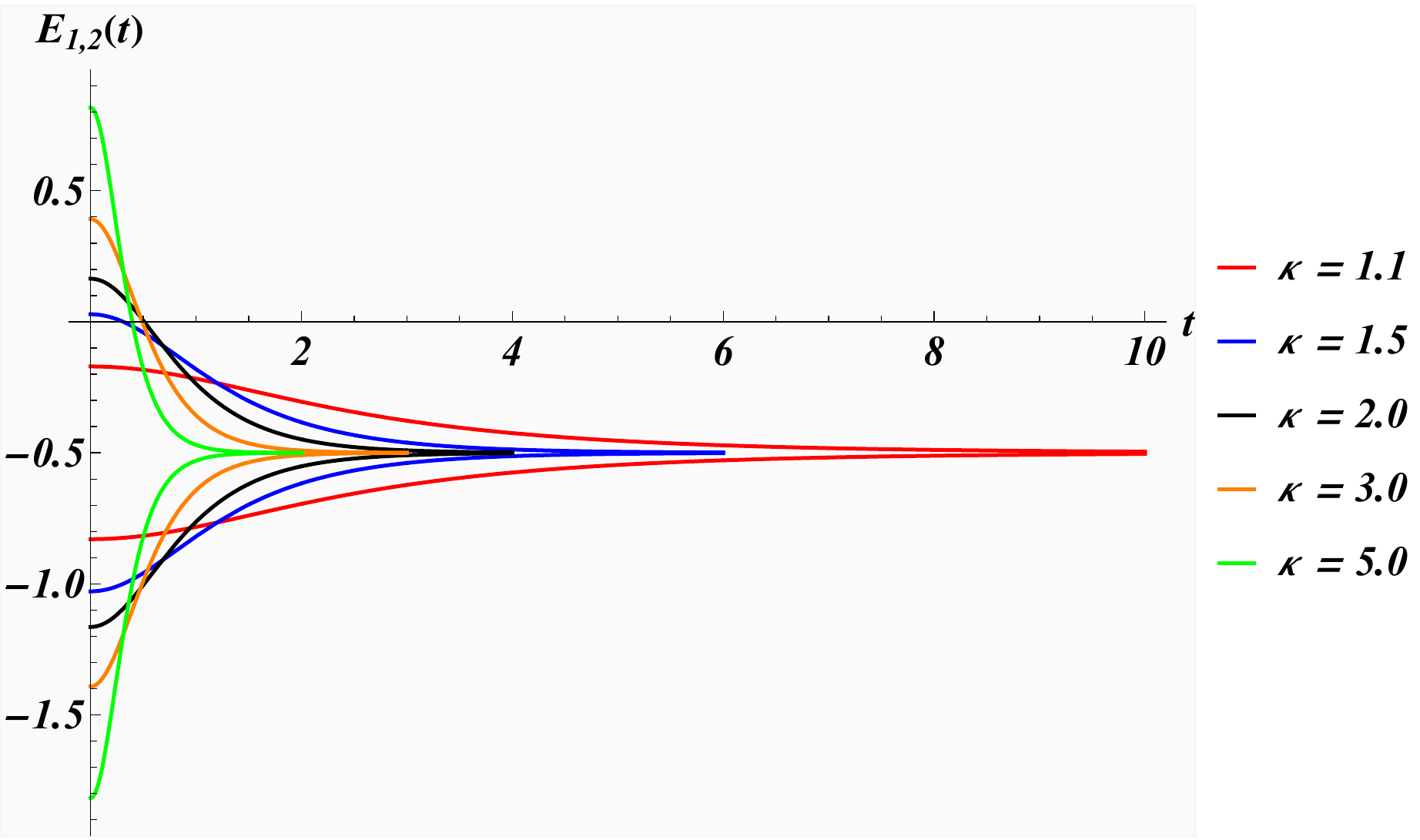}    
	\end{minipage}
	\caption{Panel (a): Time-dependent energy spectrum in what would be the $\cal{PT}$-symmetric regime in the time-independent case with $\omega=\kappa=1$ for different values of $\lambda$. Panel (b): Time-dependent energy spectrum in what would be the spontaneously broken $\cal{PT}$-regime in the time-independent case with $\omega=\lambda=1$ for different values of $\kappa$.}
	\label{enersp}
\end{figure}

For $\vert \lambda \vert > \vert \kappa \vert $ a convenient choice that leads to $\det[\rho(t)] = 1$ is $c_1=c_2=1/2$, $c_3=  \sqrt{2} \lambda/ \sqrt{\lambda^2-\kappa^2}$, $c_4=0$. Assembling all quantities, the time-dependent metric operator then takes on the form
\begin{equation}
	\rho(t)= \left(
	\begin{array}{cc}
		\frac{\kappa \sin (\zeta  t)+\sqrt{2} \lambda}{\zeta } & \cos (\zeta  t)+i \frac{ \sqrt{2} \kappa+\lambda \sin
			(\zeta  t) }{\zeta } \\
		\cos (\zeta  t)-i \frac{ \sqrt{2} \kappa+\lambda \sin (\zeta  t)}{\zeta } & \frac{\kappa \sin
			(\zeta  t)+\sqrt{2} \lambda}{\zeta } \\
	\end{array}
	\right) ,
\end{equation}
where we abbreviated $\zeta:=\sqrt{\lambda^2-\kappa^2}$. Diagonalising $\rho(t)$ we observe that the eigenvalues $\rho_\pm$ are indeed positive at all times
	\begin{equation}
		\rho(t)=UDU^{-1}, \qquad U= \left(
		\begin{array}{cc}
			iu_1 & u_1 \\
			u_2 & i u_2 \\
		\end{array}
		\right) \qquad D=\left(
		\begin{array}{cc}
			\rho_+ &0 \\
			0 & i \rho_- \\
		\end{array}
		\right),
	\end{equation}
with
\begin{equation}
u_1= \sqrt{2} \kappa +\lambda  \sin (\zeta  t)-i \zeta  \cos (\zeta  t), 
\,\, u_2=\!\! \left[\left(\sqrt{2} \lambda +\kappa  \sin (\zeta  t)\right)^2-\zeta ^2\right]^{\frac{1}{2}} \!\!, \,\, \rho_\pm= \frac{\sqrt{2} \lambda +\kappa  \sin (\zeta  t) \pm u_2}{\zeta }.
\end{equation}
Assuming that $\eta(t)$ is Hermitian, we can obtain it by taking the positive square root of the metric operator
\begin{equation}
  \eta(t) =\sqrt{\rho(t)}= U \sqrt{D} U^{-1}= \frac{1}{2}	\left(
	\begin{array}{cc}
		 \sqrt{\rho _-}+\sqrt{\rho _+} & -i\frac{ \left(\sqrt{\rho _-}-\sqrt{\rho _+}\right) u_1}{2
		 	 u_2} \\
		i\frac{ \left(\sqrt{\rho _-}-\sqrt{\rho _+}\right) u_2}{2
			 u_1} &  \sqrt{\rho _-}+\sqrt{\rho _+} \\
	\end{array}
	\right) .
\end{equation}
With explicit expressions for $\rho$ and $\eta$ at hand, we can now calculate all interesting physical quantities in the model, such as for instance the time-dependent energy expectation values at any given time for the operator as defined in (\ref{energyop}). From the expressions for $\eta(t)$ and $H$ it is straightforward, although lengthy, to calculate $\tilde{H}(t)$. Since we have kept $H$ time-independent we may still use the eigenstates $\vert \Psi_\pm \rangle$ from the time-independent scenario (\ref{eigenv}) and compute the expectation value 
\begin{equation}
	E_\pm(t):= \langle \Psi_\pm \vert \rho(t) \tilde{H}(t) \Psi_\pm \rangle . \label{timeepx}
\end{equation}
We will not present the analytical result here, but simply show the graph for $E_\pm(t) $ in figure \ref{enersp} panel a), which demonstrates that in what would be the $\cal{PT}$-symmetric regime in the time-independent case the time-dependent energies are indeed real and oscillate in time.  

Remarkably, also for $\vert \lambda \vert < \vert \kappa \vert $, which would be the spontaneously broken $\cal{PT}$-regime in the time-independent scenario, a convenient choice for the integration constants can be found so that the time-dependent metric operator becomes positive definite. Taking $c_1=-c_2=1/\sqrt{2}$, $c_3= - \lambda/ \sqrt{\kappa^2-\lambda^2}$, $c_4=0$ gives once more $\det[\rho(t)]=1$ and the time-dependent metric operator acquires the form
\begin{equation}
	\rho(t)= \left(
	\begin{array}{cc}
		\frac{\sqrt{2} \kappa  \cosh (\xi  t)-\lambda }{\xi } & \frac{-i \kappa +i \sqrt{2} \lambda  \cosh (\xi 
			t)+\sqrt{2} \xi  \sinh (\xi  t)}{\xi } \\
		\frac{i \kappa -i \sqrt{2} \lambda  \cosh (\xi  t)+\sqrt{2} \xi  \sinh (\xi  t)}{\xi } & \frac{\sqrt{2} \kappa 
			\cosh (\xi  t)-\lambda }{\xi } \\
	\end{array}
	\right) ,
\end{equation}
where we abbreviated $\xi:=\sqrt{\kappa^2-\lambda^2}$. We can now proceed in the same manner as for the complementary regime, i.e. compute the time-dependent Dyson map from the square root of $\rho(t)$ and subsequently determine $\tilde{H}(t)$. We will not present the analytical computation for this case either, but simply show the numerical evaluation in figure \ref{enersp} panel b). In this case the expectations values $E_\pm(t)$ are complex. However, we may also consider the instantaneous eigenvalue spectrum by the eigenvalue equation 
\begin{equation}
  \tilde{H}(t) \vert \tilde{\Psi}_\pm \rangle = 	E_{1,2}(t) \vert \tilde{\Psi}_\pm \rangle . \label{timeinst}
\end{equation}
and observe, see figure \ref{enersp} panel b), that $E_{1,2}(t) $ are real even 
in the what would be the spontaneously broken $\cal{PT}$-regime in the time-independent scenario.  We observe that for large time $t$ the two eigenvalues $E_{1,2}(t)$ of $\tilde{H}$ tend to the real part of the eigenvalues of $H$.

Intriguingly, in the fully time-dependent case (iii) also the energy expectation values (\ref{timeepx}) become real as we will see below. Thus in the time-dependent scenario the spontaneously broken $\cal{PT}$-regime can be made physically meaningful. We refer to this phenomenon as {\em mending the broken $\cal{PT}$-regime} \cite{AndTom3}. 

	\subsubsection{Exact time-dependent solutions for an infinite dimensional Hilbert space, example I} Next we consider the time-dependent version of the model discussed previously in section \ref{infnotime}. Following largely \cite{AndTom4} in this section, we take the Hamiltonian be explicitly time-dependent of the form
		\begin{equation}
		H_{K}(t)=a(t)\left( K_{1}+K_{2} \right)+i\lambda(t) K_{3},  \qquad a(t),\lambda(t) \in \mathbb{R}, \label{HKham}
	\end{equation}
	and have also taken $a=b$ when compared with (\ref{HK}). This would place the model strictly into the broken $\cal{PT}$-regime, but as we have seen in the previous section, the explicit time-dependence may fix the apparent issues arising from this. Instead of solving the TDQHE (\ref{TDQHE}) as in the last section, we will solve now instead the TDDE (\ref{TDDE}). We exploit once more the fact that the Hamiltonian is expanded in terms of generators of a closed algebra (\ref{alg}). Making a similar general Ansatz as in section \ref{infnotime}
	\begin{equation}
		\eta (t)=\dprod\nolimits_{i=1}^{4}e^{\gamma _{i}(t)K_{i}},\qquad \gamma
		_{i}(t)\in \mathbb{R}, \label{ansatzeta}
	\end{equation} 
	but now with time-dependent coefficient functions $\gamma_{i}\rightarrow \gamma_{i}(t)$, its substitution into the TDDE will only produce terms that are also in this algebra. This is clear for the adjoint action of $\eta$ on $H$ as we have already seen and also for the second gauge like term as it is a Maurer-Cartan form that is an element of the algebra, see e.g. \cite{helgason1979diff}. Thus with the Ansatz (\ref{ansatzeta}) the TDDE (\ref{TDDE}) is satisfied when the time-dependent coefficient functions $\gamma_{i}(t)$ constrained as
	\begin{equation}
		\gamma _{1}=c_{1},\quad \gamma _{2}=c_{2}, \quad \dot{\gamma}_{3}=-\lambda \cosh \gamma
		_{4},\quad \dot{\gamma}_{4}=\lambda \tanh \gamma _{3}\sinh \gamma _{4},
		\label{34}
	\end{equation}
with $c_{1},c_{2}$ some real constants, and the time-dependent Hermitian Hamiltonian resulting to
	\begin{equation}
		h(t)=a(t)\left( K_{1}+K_{2}\right) +\frac{\lambda (t)}{2}\frac{\sinh \gamma
			_{4}}{\cosh \gamma _{3}}\left( K_{1}-K_{2}\right) .
	\end{equation} 
We still need to solve the last two equations in (\ref{34}), which can be achieved by dividing them by each other, hence eliminating the differential in time and $\lambda$, separation of variables and a subsequent integration
	\begin{equation}
  \int \frac{1}{\tanh \gamma _{4}}	d\gamma_{4}= - \int  \tanh \gamma _{3} d\gamma_{3} \qquad \Rightarrow \quad \gamma _{4}=\func{arcsinh}\left( c_{3} \func{sech}\gamma _{3}\right), \label{g34}
\end{equation}
with integration constant $c_{3}$. Trading now in the last two equations in (\ref{34}) $\gamma
_{4}$ for $\gamma_{3}$, with the help of (\ref{g34}), and parametrizing $\gamma_{3} = \arccosh \left[ \chi(t) \right]$ converts these two equations into a dissipative version of the nonlinear Ermakov-Pinney (EP) equation \cite{Ermakov,Pinney}
\begin{equation}
	\ddot{\chi}-\frac{\dot{\lambda}}{\lambda }\dot{\chi}-\lambda ^{2}\chi =\frac{%
			c_{3} ^{2}\lambda ^{2}}{\chi ^{3}} . \label{EP1}
\end{equation} 
It is quite common in the context of time-dependent systems, that their solutions are parametrised by solutions of different variants of the EP-equation. As the EP-equation is nonlinear, the solutions procedures are not standard and not all versions have been solved analytically. The original form of this equation was solved by Pinney \cite{Pinney}, see (\ref{solPinney}), and solutions to various other systems are discussed for instance in \cite{PhysRevD.90.084005,dey2015milne,AndTom1,AndTom3,AndTom2,AndTom4,cen2019time,BeckyAnd1,fring2021perturb,fring2021exactly}. The solution to (\ref{EP1}) constructed in \cite{AndTom4} acquires the form 
\begin{equation}
 \chi(t)=	\sqrt{\left(1+c_{3} ^2\right) \cosh ^2 \left[c_4-\int^t \lambda (s) \, ds\right]-c_{3} ^2} . \label{solEP}
\end{equation}
Note that there is only one additions integration constant $c_4$, despite the fact that we are solving a second order differential equation, due to the fact one constant, $c_3$, was already introduced previously.

We may now compute all physical quantities of interest, notable the energy operator as introduced in equation (\ref{energyop})
\begin{equation}
	\tilde{H}(t)=a(t)\left( K_{1}+K_{2}\right) +\frac{\lambda (t)}{4}\sinh
	(2\gamma _{4})\left( K_{1}-K_{2}\right) -i\lambda (t)\left( \sinh ^{2}\gamma
	_{4}K_{3}-\sinh \gamma _{4}\tanh \gamma _{3}K_{4}\right) . \label{97}
\end{equation}
All quantities in (\ref{97}) are known, with $a(t), \lambda(t)$ specified in $H(t)$, $\gamma_3(\chi)$, $\gamma_4(\gamma_3)$ and $\chi$ given in (\ref{solEP}). Time-dependent expectation values for this system are discussed in \cite{fring2021perturb}. The remarkable feature is that the expectation value for $\tilde{H}(t)$ is real, so that the broken $\cal{PT}$-regime has been mended. 

		\subsubsection{Exact time-dependent solutions for an infinite dimensional Hilbert space, example II} \label{anharmsec}
\begin{figure}[h]
	\noindent	\begin{minipage}[b]{0.54\textwidth}     \!\!\!\! \!\!\!\! \includegraphics[width=\textwidth]{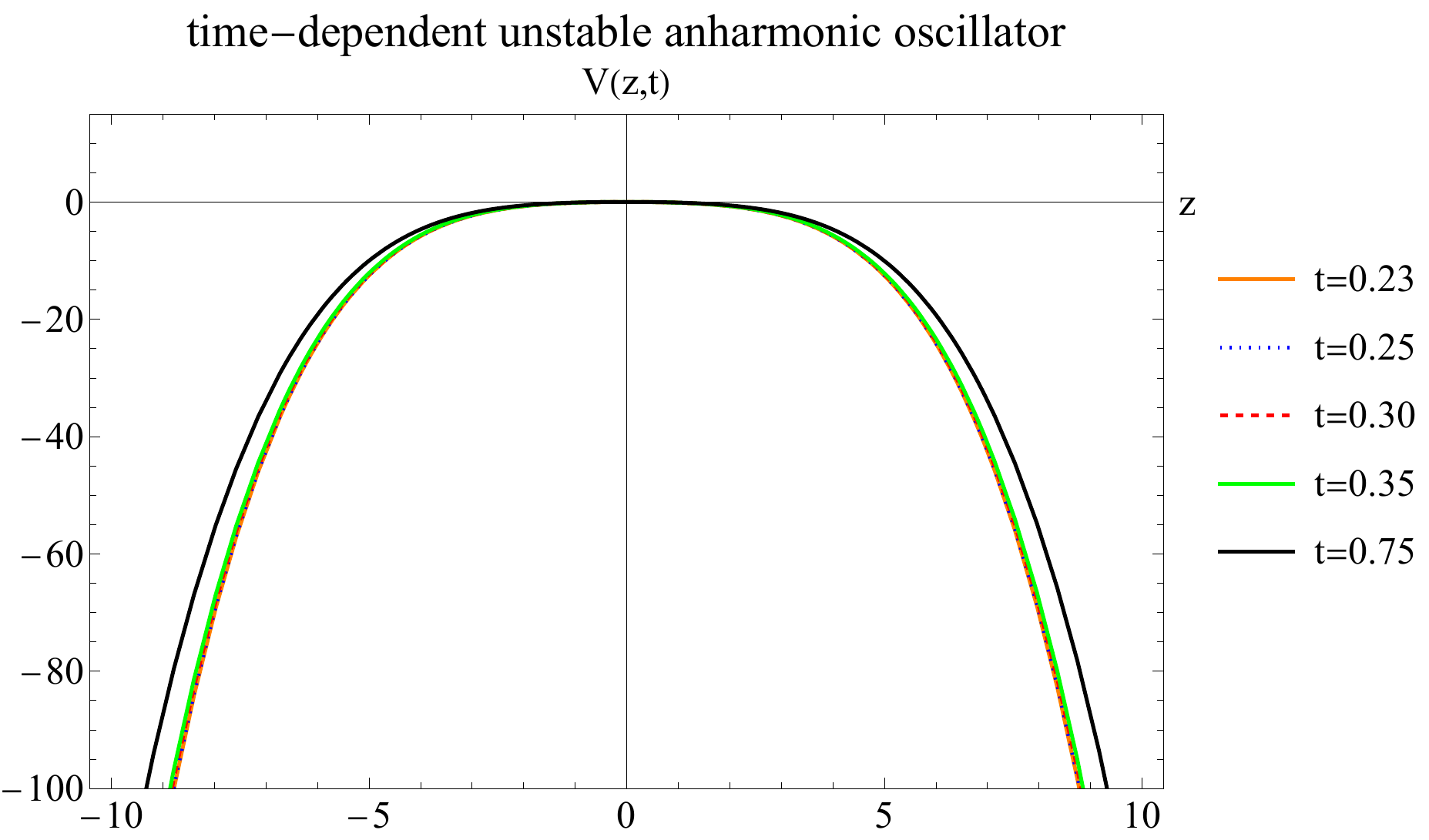}
	\end{minipage}
	\begin{minipage}[b]{0.46\textwidth}      \includegraphics[width=\textwidth]{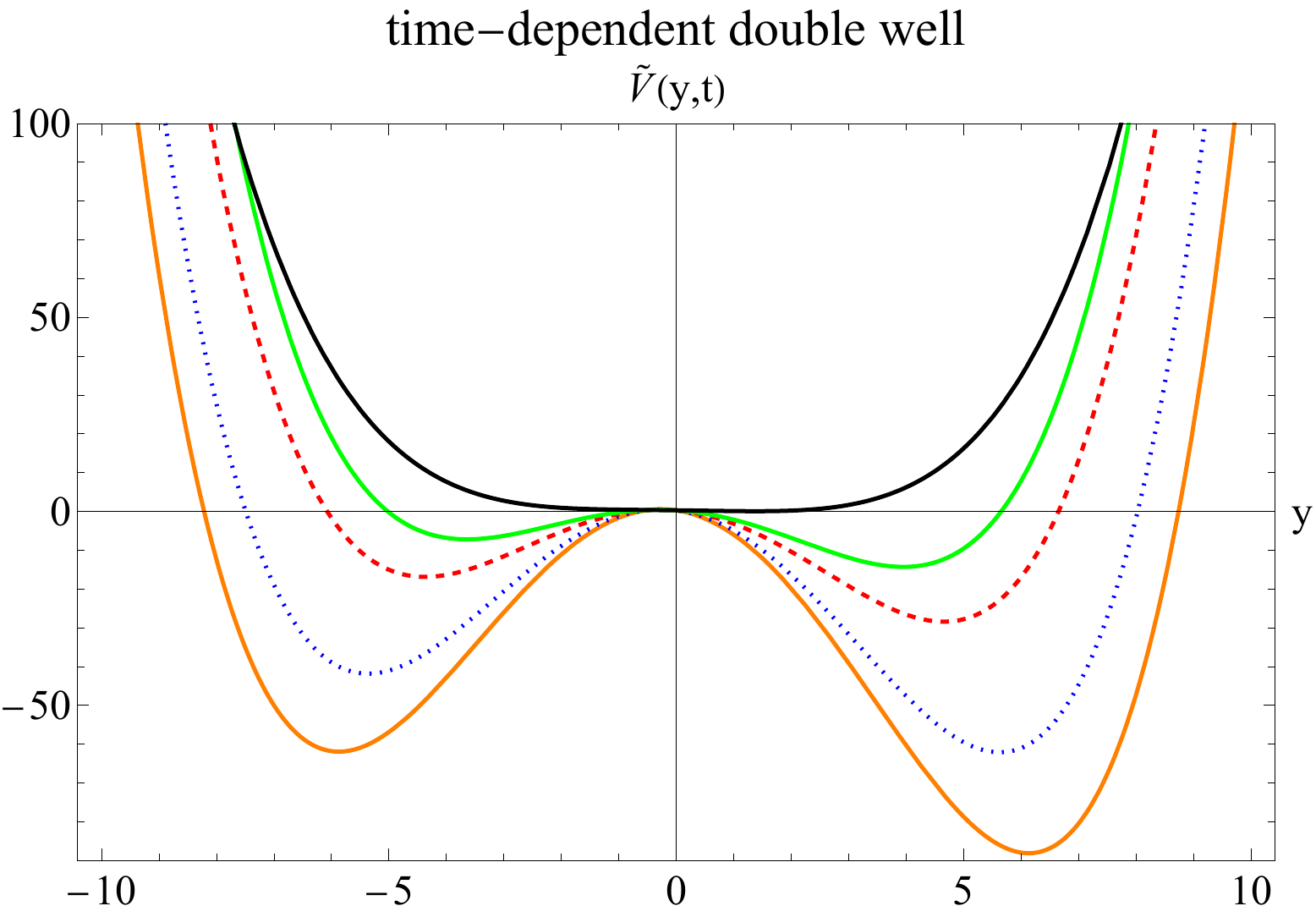}    
	\end{minipage}
	\caption{Spectrally
		equivalent time-dependent anharmonic oscillator potential $V(z,t)$ in (\ref{H4}) and time-dependent double well potential $\tilde{V}(y,t)$ in (\ref{doublewell}) for 
		$\sigma (t)=\cosh t$, $g(t)=1/4\cosh ^{3}t$, $m(t)=(\tanh ^{2}t-2)/4$ at
		different values of time.}
	\label{figdouble}
\end{figure}

		The unstable harmonic oscillator has been a classic example of an exactly solvable system in the time-independent scenario \cite{JM}, in the sense that one can find an exact analytical expression for the Dyson map. Apart from the additional confining mass term mass, this model can be seen as the $\varepsilon=2$-case of the Bender-Boettcher potential in (\ref{HBB}). Despite the potential being real, we would still anticipate this to be an ill-defined model from a conventional point of view as the potential is unbounded from below and hence is expected not to possess a proper ground state. Here we present the discussion from \cite{BeckyAnd2} for the time-dependent version defined by the Hamiltonian
	\begin{equation}
		H_4(z,t)=p^{2}+\frac{m(t)}{4}z^{2}-\frac{g(t)}{16}z^{4},~~~~~m(t)\in \mathbb{R}%
		\text{,}g(t)\in \mathbb{R}^{+} . \label{H4}
	\end{equation}
	Following Jones and Mateo \cite{JM} we regard $z$ to be complex and define model on the contour  $z=-2i\sqrt{1+ix}$, which leads to   
	\begin{equation}
		H_4(x,t)=p^{2}-\frac{1}{2}p+\frac{i}{2}\{x,p^{2}\}-m(t)(1+ix)+g(t)(x-i)^{2} .
	\end{equation} 
As this Hamiltonian is not expressed in terms of generators of a closed algebra, it is more difficult to make a definite Ansatz for the Dyson map. One may of course attempt using trial and error by including terms in the exponent that are present in the Hamiltonian and those that are generated by multiple nested commutators as present in the BCH-formula. However, as new terms keep being generated when expanding further this becomes rather complex and it seems to be very surprising to find a closed formula at all. In fact in the time-independent case the closed solution for $\eta$ was found systematically by using perturbation theory as explained in general in section \ref{general}. Here we present the Ansatz 
	\begin{equation}
	\eta (t)=e^{\alpha (t)x}e^{\beta (t)p^{3}+i\gamma (t)p^{2}+i\delta
		(t)p},~~~~~\alpha ,\beta ,\gamma ,\delta \in \mathbb{R} . \label{ansatzeta4}
\end{equation}
that leads to an exact solution, which may also be obtained from generalization to time-dependent perturbation theory as will be discussed in section \ref{pertsec}. It is not obvious that this will indeed lead to a solution and moreover we have also taken some of the coefficient functions to be purely imaginary, which seems to be unmotivated. Inserting (\ref{ansatzeta4}) into the TDDE (\ref{TDDE}), we only need to repeatedly use the canonical commutation relations $[x,p]=1$ in the BCH formula and find that the coefficient functions $\alpha(t) ,\beta(t) ,\gamma(t) ,\delta(t) $ have to be constrained to
	\begin{equation}
		\alpha =\frac{\dot{g}}{6g},~~~\beta =\frac{1}{6g},~~~\gamma =\frac{%
			12g^{3}+6mg^{2}+\dot{g}^{2}-g\ddot{g}}{4\dot{g}g^{2}},~~~\delta =c_{1}\frac{g%
		}{\dot{g}}-\frac{g\ln g}{2\dot{g}},  \label{const}
	\end{equation}
where a new function $g(t)$ has been introduced to parametrize these function and needs to satisfy the third order differential equation
\begin{equation}
	9g^{2}\left( \dddot{g}-6g\dot{m}\right) +36g\dot{g}\left( gm-\ddot{g}\right)
	+28\dot{g}^{3}=0 .  \label{thirdo}
\end{equation}
The time-dependent Hermitian counterpart Hamiltonian on the LHS of (\ref{TDDE}) takes on the form  
\begin{eqnarray}
				h_4(x,t) \!\! &=& \!\!\frac{p^{4}}{4g}+\left[ \frac{18g^{2}(2g+m)}{\dot{g}^{2}}+\frac{%
				\dot{g}^{2}}{72g^{3}}-\frac{2g+m}{4g}\right] p^{2}-\frac{3\left(
				g^{2}m+g^{3}\right) \ln g}{\dot{g}^{2}}p+\frac{g^{2}\ln (g)}{\dot{g}}x~~  \qquad \\
			&& +\left( \frac{\dot{g}}{12g}-\frac{6g^{2}}{\dot{g}}\right) \text{$\{x,p\}$}%
			+gx^{2}+\frac{1296g^{8}\ln ^{2}g+\dot{g}^{6}-36\dot{g}^{4}g^{2}(2g+m)}{%
				5184g^{5}\dot{g}^{2}}-\frac{m}{2} .\notag
	\end{eqnarray}
Parametrising $g(t),m(t)$ by means of a new function $\sigma(t)$
\begin{equation}
	g=\frac{1}{4\sigma ^{3}},\quad ~~\text{and\quad ~~}m=\frac{4c_{2}+\dot{\sigma%
		}^{2}-2\sigma \ddot{\sigma}}{4\sigma ^{2}}, \label{mgg}
\end{equation}
equation (\ref{thirdo}) is solved with free function $\sigma(t)$ and constant $c_1$. Since $m(t)$ is given in the Hamiltonian, we may determine $\sigma$ from the last equation in (\ref{mgg}) for a specific $m(t)$, provided such a solution exists. It is useful to analyse the Hermitian Hamiltonian further 
by unitary transforming it, scaling it and subsequently Fourier transform it, ending up with the Hamiltonian 
\begin{eqnarray}
	\tilde{h}_4(y,t)  \!\! &=&  \!\! p_y^2 +  \frac{g}{4}y^{2}\left[ y^{2}+\frac{\dot{g}^{2}}{36g^{3}}+\frac{72g^{2}m}{%
		\dot{g}^{2}}-\frac{m}{g}+2\right] +\frac{\left( 36g^{2}m+\dot{g}^{2}\right) 
		\sqrt{g}\ln g}{12\dot{g}^{2}}y \label{doublewell} \\
	&& +\frac{\dot{g}^{4}}{5184g^{5}}-\frac{\dot{g}^{2}m}{144g^{3}}-\frac{\dot{g}%
		^{2}}{72g^{2}}-\frac{m}{2} , \notag
\end{eqnarray} 
involving a time-dependent double well potential. We refer the reader to \cite{BeckyAnd2} for further details on these calculations. In summary, we have carried out the following manipulations
	\begin{equation}
		H_4(z,t)\overset{z\rightarrow x}{\rightarrow }H_4(x,t)\overset{\text{Dyson}}{%
			\rightarrow }h_4(x,t)\overset{\text{unitary transform}}{\rightarrow }\hat{h}_4%
		(x,t)\overset{\text{Fourier}}{\rightarrow }\tilde{h}_4(y,t),
\end{equation}
and overall have converted the time-dependent unstable anharmonic oscillator into a  time-dependent double well potential as illustrated in figure \ref{figdouble} 	

		\subsubsection{Perturbative approach} \label{pertsec}
			As already indicated, it will often be a matter of guesswork to find a suitable Ansatz for $\eta(t)$ or $\rho(t)$. Thus to enable a more systematic construction it is therefore useful to generalise the time-independent perturbation theory as discussed in section \ref{general} to the time-dependent case. As it turn out here we need to distinguish between two possibilities, a weakly and a strongly coupled non-Hermitian term or terms \cite{fring2021perturb}. \\
			\underline{Weakly coupled systems:}\\ We start by decomposing our time-dependent non-Hermitian Hamiltonian as
		\begin{equation}
			H(t)=h_{0}(t)+i\epsilon h_{1}(t),\qquad \ \ \text{with}~h_{0}(t)=h_{0}^{%
				\dagger }(t),h_{1}(t)=h_{1}^{\dagger }(t), \epsilon \ll 1,
		\end{equation}
	whereby $\epsilon$ is small, which is what we mean by weakly coupled. Attempting now to solve the TDQHE (\ref{TDQHE}) perturbatively, we take the metric to be of the form $\rho(t)=\exp(q(t))$. Comparing with the expressions in section \ref{general}, suggest to replace in the first instance
		\begin{equation}
		q=\sum_{n=1}^{\infty }\epsilon ^{n}\check{q}_{n} \quad \rightarrow \quad	q(t)=2\sum_{n=1}^{\infty }\sum_{i=1}^{N_{n}}\epsilon ^{n}\tilde{\gamma}%
			_{i}^{(n)}(t)\tilde{q}_{i}^{(n)} . \label{108}
		\end{equation}
	We have taken here into account that at each order we may have a number of terms, $N_n$, labelled by the index $i$ with time-dependent coefficient functions $\tilde{\gamma} _{i}^{(n)}(t)$ and operators $\tilde{q}_{i}^{(n)}$. The factor 2 is only convenience. However, this form $e^{\tilde{A}(t)+\tilde{B}(t)+\tilde{C}(t)+\ldots }$ with $\tilde{A}(t),\tilde{B}(t),\tilde{C}(t)\ldots$ being in general non-commuting operators is difficult to differentiate with respect to $t$. We prefer a factorised form $e^{A(t)}e^{B(t)}e^{C(t)}\ldots$, which is easily differentiated  $\partial_t(e^{A(t)}e^{B(t)}e^{C(t)}\ldots) = \dot{A}(t)e^{A(t)}e^{B(t)}e^{C(t)} + e^{A(t)}\dot{B}(t)e^{B(t)}e^{C(t)}++ e^{A(t)}e^{B(t)}\dot{C}(t)e^{C(t)}+ \ldots$. The conversion from the operators $\tilde{A}(t),\tilde{B}(t),\tilde{C}(t)\ldots$ to $A(t),B(t),C(t)\ldots$ is in general nontrivial, but can also be omitted as we can build in the factorisation directly into our Ansatz for the perturbation theory. We therefore specify (\ref{108}) further into 
		\begin{equation}
			q(t)=2\sum_{n=1}^{\infty}\sum_{i=1}^{j}\epsilon ^{n}\gamma _{i}^{(n)}(t)q_{i}=2\sum_{i=1}^{j}\sum_{n=1}^{k}\epsilon ^{n}\gamma _{i}^{(n)}(t)q_{i}.
		\end{equation} 
	We assumed here that $\tilde{q}_{i}^{(n)} \in \{q_1,q_2, \dots q_j\}$ so that we can drop the explicit mentioning of the order $n$. In the second step we swapped the two sums and also terminate the infinite sum at a finite value $k$. Taking the $q_i$ to be Hermitian, the metric then acquired the general form
		\begin{equation}
			\rho (t)=\eta (t)^{\dagger }\eta (t)=\prod_{i=j}^{1}\left[
			\prod_{n=k}^{1}\exp \left( \epsilon ^{n}\gamma _{i}^{(n)}q_{i}\right) \right]
			\prod_{i=1}^{j}\left[ \prod_{n=1}^{k}\exp \left( \epsilon ^{n}\gamma
			_{i}^{(n)}q_{i}\right) \right] 
		\end{equation}
		where we understand the products as being ordered, that is $\prod_{i=1}^{j}a_{i}=a_{1}a_{2}\ldots a_{j}$ whereas $\prod_{i=j}^{1}a_{i}=a_{j}a_{j-1}\ldots a_{1}$. Setting now $k=1$ equation (\ref{TDQHE}) becomes
			\begin{equation}
			ih_{1}+\sum_{i=1}^{j}\left( \gamma _{i}^{(1)}\left[ q_{i},h_{0}\right] +i%
			\dot{\gamma}_{i}^{(1)}q_{i}\right) =0,
		\end{equation}
	and for $k=2$ the second order equation reads 
	 \begin{eqnarray}
			0 \!\!	&=&\!\! 2\sum_{i=1}^{j}\left( \gamma _{i}^{(2)}[q_{i},h_{0}]+i\gamma
				_{i}^{(1)}[q_{i}^{1},h_{1}]+\frac{1}{2!}(\gamma
				_{i}^{(1)})^{2}[q_{i},[q_{i},h_{0}]]+i\dot{\gamma}_{i}^{(2)}q_{i}\right) 
				\\
				&& +\sum_{i=1}^{j}\left( 2\sum_{r=1,\neq i}^{j}\left( \gamma _{i}^{(1)}\gamma
				_{r}^{(1)}[q_{r},[q_{i},h_{0}]]+i\dot{\gamma}_{i}^{(1)}\gamma
				_{r}^{(1)}[q_{r},q_{i}]\right) +(\gamma
				_{i}^{(1)})^{2}[q_{i},[q_{i},h_{0}]]\right) . \notag
		\end{eqnarray}
Despite the fact that these equations are more complicated due to the additional differential of the coefficient functions, they can be solved recursively as in the time-independent case. However, not all systems can be treated in this manner, such as the example in section \ref{anharmsec} and we also require a version of perturbation theory suitable for strongly coupled systems.\\
\underline{Strongly coupled systems:} \\ We now separate the time-dependent non-Hermitian Hamiltonian into three terms
\begin{equation}
	H(t)=h_{1}(t)+\epsilon ^{2}h_{2}(t)+i\epsilon h_{3}(t)\qquad \ \ \text{with}~h_{i}(t)=h_{i}^{%
		\dagger }(t), i=1,2,3, \, \epsilon \gg 1,
\end{equation}
with the key difference being that $\epsilon$ is taken to be very large, which is what we mean by strongly coupled. The following Ansatz 
\begin{equation}
	\rho (t)=\eta (t)^{\dagger }\eta (t)=\prod_{i=j}^{1}\left[
	\prod_{l=k}^{1}\exp \left( \epsilon ^{-l}(\gamma _{i}^{(l)})^{\dagger
	}q_{i}\right) \right] \prod_{i=1}^{j}\left[ \prod_{l=1}^{k}\exp \left(
	\epsilon ^{-l}(\gamma _{i}^{(l)})q_{i}\right) \right] ,
\end{equation}
is then motivated in a similar way as for the weakly coupled system, with the difference that we need to take the inverse of $\epsilon$. Once more the set of equations we obtain order by order can be solved recursively. Using this expansion we have shown in \cite{fring2021perturb} that the exact solution presented in section \ref{anharmsec} can indeed be recovered. 
		
		\subsubsection{Utilizing Lewis-Riesenfeld invariants} \label{LRinv}
		A very useful method to find exact solutions to the TDSE for time-dependent systems consists of employing Lewis-Riesenfeld invariants \cite{LewisR69}. As the name suggests, the central objects involved are conserved quantities satisfying the conservation equation 
		 \begin{equation}
		 	\frac{dI_{\mathcal{H}}(t)}{dt} = \partial _{t}I_{\mathcal{H}}(t)-\frac{i}{\hbar} \left[ I_{\mathcal{H}}(t),\mathcal{H}%
		 	(t)\right]=0 ,\quad   \text{for~\ }\mathcal{H}=h=h^{\dagger } \,\, \text{or} \,\, \mathcal{H}=H\neq	H^{\dagger } . \label{LRinvdef}
		 \end{equation} 
	 As indicated, we may consider this equation for Hermitian or non-Hermitian Hamiltonians. The aim is to solve the TDSE $\mathcal{H}(t)\Psi_{\mathcal{H}} (t) =i\hbar \partial _{t}\Psi_{\mathcal{H}} (t)$. As shown in much detail in \cite{LewisR69,PhysRevD.90.084005}, the invariants satisfy the following key properties
		 \begin{eqnarray}
		 	~~~I_{\mathcal{H}}(t)\left\vert \phi _{\mathcal{H}}(t)\right\rangle
		 \!\!	&=& \!\!\Lambda \left\vert \phi _{\mathcal{H}}(t)\right\rangle ,~~~~~~~
		 	\ \ ~\ ~ \quad  \dot{\Lambda}=0  , \label{LR1}
		 	\\
		 	 \left\vert \Psi _{\mathcal{H}}(t)\right\rangle \!\! &=& \!\! e^{i\hbar
		 		\alpha (t)}\left\vert \phi _{\mathcal{H}}(t)\right\rangle 
		 	,\qquad \quad   \dot{\alpha}= \left\langle \phi _{\mathcal{H}}(t)\right\vert i\hbar
		 	\partial _{t}-\mathcal{H}(t)\left\vert \phi _{\mathcal{H}}(t)\right\rangle  . \label{LR2}
		 \end{eqnarray}  
	 The first equation in (\ref{LR1}) is the crux of the method, as it means that one has reduced the TDSE to a much simpler eigenvalue problem for the invariant in which time is simply a parameter as any other. If one is able to solve this eigenvalue equation, one obtains the solution to the TDSE by a phase factor which can be determined from the second equation in (\ref{LR2}). Below we will also comment on how to proceed in case one does not succeed in solving the eigevalue equation. 
	 
		 The key property we shall be exploiting here is the fact that the non-Hermitian invariant $I_{H}$ is quasi-Hermitian
		 \begin{equation}
		 	I_{h}(t) = \eta (t)I_{H}(t)\eta ^{-1}(t), \label{invquasi}
  	\end{equation}
  noting that $I_{h}$ is necessarily Hermitian as follows directly from the defining relation (\ref{LRinvdef}). We can easily prove (\ref{invquasi}). Starting by assuming it to hold we differentiate it with respect to $t$ and subsequently by using (\ref{invquasi}) for $H$ and (\ref{TDDE}), we derive (\ref{invquasi}) for $h$
  \begin{eqnarray}
  	\dot{I}_{h} \!\!&=&\!\! \dot{\eta} I_{H}\eta ^{-1} + \eta \dot{I}_{H}\eta ^{-1} + \eta I_{H}\dot{\eta} ^{-1}\\
  	 &=&\!\! \dot{\eta} I_{H}\eta ^{-1} + i \eta \left[ I_{H} ,H \right] \eta^{-1} - \eta \dot{I}_{H} \eta^{-1} \dot{\eta} \eta^{-1} \notag \\
  	 &=&\!\! i \left( -i  \dot{\eta} \eta^{-1} \eta I_{H}\eta ^{-1} - \eta H \eta^{-1} \eta I_{H} \eta^{-1} + i \eta I_{H}\eta ^{-1} \dot{\eta} \eta^{-1} +
  	    \eta I_{H}\eta ^{-1} \eta H \eta^{-1}  \right) \notag \\
  	     &=&\!\! i \left[  \eta I_H \eta ^{-1}, \eta H\eta ^{-1} + i  \dot{\eta} \eta^{-1}   \right] \notag \\
  	     &=&\!\!i\left[  I_h, h  \right] . \notag
  \end{eqnarray}
  Thus we have now obtained the possibility to pursue a new procedure: Constructing first the two invariants $I_{h}(t)$ and $I_{H}(t)$ we can try to find $\eta(t)$ from the quasi-Hermiticity relation (\ref{invquasi}). This means we have translated the problem of solving the time-dependent Dyson equation (\ref{TDDE}) to the far simpler problem of find a similarity transformation for the invariants in which time just plays the role of a parameter. This is now akin to the time-independent case with Hamiltonians replaced by invariants. Of course this comes at the cost of actually having to construct the invariants. Let us now see how to obtain them. 
  
  The standard way to construct invariants consists of making an Ansatz for them by including terms that are already present in the time-dependent Hamiltonian and possibly others that results from the commutation relations of these terms. Despite of this being a bit of guesswork many exact invariants have been constructed in this manner, for Hermitian and non-Hermitian systems. The procedure to find $\eta$ from (\ref{invquasi}) has been applied successfully in \cite{maamache2017pseudo,khantoul2017invariant,AndTom4,cen2019time,fring2021exactly}. 
    \\  
		  \underline{Exact invariants and exact solutions from point transformations} \\
Alternatively one can make use of the fact that point transformations preserve conserved quantities, see e.g. \cite{steeb1993invertible}. This was first shown to be applicable in the construction of Lewis-Riesenfeld invariants for time-dependent Hermitian Hamiltonian systems in \cite{zelaya2020quantum} and generalised to case of time-dependent non-Hermitian Hamiltonian systems in \cite{fring2021exactly}.

 The starting point of the construction is the TDSE 
	\begin{equation}
		H_{0}(\chi )\psi (\chi ,\tau )=i\hbar \partial _{\tau }\psi (\chi ,\tau ) , \label{refH}
	\end{equation}
for what we refer to as the {\em reference Hamitonian} $H_{0}(\chi )$ depending only on the coordinate $\chi$ and not on the time $\tau$. One then seeks a transformation $\Gamma$ that maps the TDSE (\ref{refH}) to the TDSE $H(x,t)\phi (x ,t )=i\hbar \partial _{t }\phi (x ,t ) $, involving the non-Hermitian explicitly time-dependent {\em target Hamiltonian} $H(x,t)$  
	\begin{equation}
		\Gamma :H_{0}\text{-TDSE}\rightarrow H\text{-TDSE, \ \ \ \ \ \ }[\chi ,\tau
		,\psi (\chi ,\tau )]\mapsto \left[ x,t,\phi (x,t)\right] . \label{gammap}
	\end{equation}
In general, the reference variables $\chi ,\tau ,\psi$ and the target variables $x,t,\phi $ are regarded as independent  
	\begin{equation}
		\chi =P(x,t,\phi ),\text{\qquad }\tau =Q(x,t,\phi ),\qquad \psi =R(x,t,\phi
		) , \label{PQR}
	\end{equation}
but as indicated in (\ref{gammap}), here $\psi $ and $\phi $ are treated as implicit functions of $(\chi $,$\tau )$ and $(x$,$t)$, respectively. The key point that enables the construction procedure in the first place is the fact that point transformations preserve invariants. Thus constructing the map $\Gamma$ in the first instance from mapping the two TDSE into each other, one may subsequently use it to act on the reference Hamiltonian alone and thus obtains an invariant for the target Hamiltonian
	\begin{equation}
		\Gamma :H_{0}(\chi )\rightarrow I_{H}(x,t) .
	\end{equation}	
We illustrate this with an example: The reference Hamiltonian is ideally chosen as a simple solvable system. Here we take it to be the time-independent Hermitian oscillator Hamiltonian
\begin{equation}
	H_{0}(\chi )=\frac{P^{2}}{2m}+\frac{1}{2}m\omega ^{2}\chi
	^{2},~~~~~~~m,\omega \in \mathbb{R} ,
\end{equation}
for which solutions can be found in any elementary book on quantum mechanics. To facilitate the computations we make some simplifying assumptions on the general dependences (\ref{PQR}) 
\begin{equation}
	\chi =\chi (x,t),\qquad \tau =\tau (t),\qquad \psi =A(x,t)\phi (x,t) .
\end{equation} 
The first two choices are made for convenience and the last factorisation is a consequence of the fact that our target Hamiltonian does not contain terms of the form $\psi_{\phi \phi} =0$. Using the standard representation for the momentum operator $P = - i \partial_\chi $ we translate all reference variables to the target variables including their differentials. Then the $H_{0}$-TDSE converts into 
\begin{equation}
	i\hbar \phi _{t}+\frac{\hbar ^{2}}{2m}\frac{\tau _{t}}{\chi _{x}^{2}}\phi
	_{xx}+B_{0}(x,t)\phi _{x}-V_{0}(x,t)\phi =0 \label{targetTDSE}
\end{equation}
with%
\begin{eqnarray}
	B_{0}(x,t) &=&-i\hbar \frac{\chi _{t}}{\chi _{x}}+\frac{\hbar ^{2}}{2m}\frac{%
		\tau _{t}}{\chi _{x}^{2}}\left( 2\frac{A_{x}}{A}-\frac{\chi _{xx}}{\chi _{x}}%
	\right) ,  \label{B0} \\
	V_{0}(x,t) &=&\frac{1}{2}m\tau _{t}\chi ^{2}\omega ^{2}-i\hbar \left( \frac{%
		A_{t}}{A}-\frac{A_{x}\chi _{t}}{A\chi _{x}}\right) -\frac{\hbar ^{2}}{2m}%
	\frac{\tau _{t}}{\chi _{x}^{2}}\left( \frac{A_{xx}}{A}-\frac{A_{x}\chi _{xx}%
	}{A\chi _{x}}\right) .
\end{eqnarray}	
Starting from different types of reference Hamiltonians will of course produce different target TDSEs. For instance, we find \cite{fring2021exactly}
\begin{eqnarray}
		H_{0}^{(1)} \!\!\!& =& \!\!\!\frac{P^{2}}{2m}    \rightarrow  \	B_{1}(x,t) =B_{0},\quad  V_{1}(x,t) =V_{0}-\frac{1}{2}m\omega
		^{2}\chi ^{2}\tau _{t},  \\
		H_{0}^{(2)}\!\!\!&=&\!\!\!H_{0}+a\chi \rightarrow B_{2}(x,t) =B_{0},\quad  V_{2}(x,t) =V_{0}+a\chi \tau _{t},\\
			H_{0}^{(3)}\!\!\!&=&\!\!\! H_{0}+ib\chi \rightarrow  	B_{3}(x,t) =B_{0},\quad  V_{3}(x,t) =V_{0}+ib\chi \tau _{t},
			\label{b3} \\
			H_{0}^{(4)}\!\!\!&=&\!\!\!H_{0}+a\{\chi ,P\} \rightarrow  B_{4}(x,t) =B_{0}+\frac{2ia\hbar \chi \tau _{t}}{\chi _{x}},\,  
		\quad	V_{4}(x,t) =V_{0}-\frac{2ia\chi \hbar A_{x}\tau _{t}}{A\chi _{x}}%
			-ia\hbar \tau _{t}, \qquad
\end{eqnarray}   
 with $a,b \in \mathbb{R}$, where the $B_{0},V_{0}$ in (\ref{targetTDSE}) are to be replaced by $B_{i},V{i}$, $i=1,2,3,4$. 
 
 Next we chose as a concrete target Hamiltonian the Swanson model \cite{Swanson}, which is a prototype non-Hermitian system for which many aspects have been studied, including its spontaneously broken ${\cal PT}$-regime \cite{Marta} and its time-dependent version \cite{fringmoussa2,fring2021exactly}. Usually the Swanson model is presented in terms of creation and annihilation operators $a,a^\dagger$, but since our point transformation acts in space and time we convert it to  
	\begin{equation}
	H_{S}(x,t):=\frac{p^{2}}{2M(t)}+\frac{%
		M(t)}{2}\Omega ^{2}(t)x^{2}+i\alpha (t)\{x,p\},~~M,\Omega \in \mathbb{R}%
	\text{, }\alpha \in \mathbb{C},
\end{equation}
by using the standard representation for the $a,a^\dagger$ in terms of $x,p$ \cite{fring2021exactly}. We also allow the mass to be explicitly time-dependent $M(t)$. As long as $\alpha \neq 0$ the Swanson Hamiltonian $H_{S}$ is non-Hermitian, but it is $\mathcal{PT}$-symmetric with $\mathcal{PT}$: $x\rightarrow -x$%
, $p\rightarrow p$, $i\rightarrow -i$ when all time-dependent coefficient
functions transform as $\mathcal{PT}$: $M,\Omega ,\alpha \rightarrow
M,\Omega ,\alpha $. By allowing $\alpha(t)$ to be complex we made use of a new option that did not exist in the time-independent case as it would break the $\mathcal{PT}$-symmetry already at the level of the Hamiltonian. However, in the time-dependent scenario we can maintain the $\mathcal{PT}$-symmetry by demanding  $\mathcal{PT}$: $\alpha _{R}\rightarrow \alpha _{R}$, 
$\alpha _{I}\rightarrow -\alpha _{I}$ for $\alpha =\alpha _{R}+i\alpha _{I}$.
For the ease of presentation we focus here first on the case of time-independent mass $M(t) \rightarrow m$. Then the $H_{S}$-TDSE takes on the form 
\begin{equation}
	i\hbar \phi _{t}+\frac{\hbar ^{2}}{2m}\phi _{xx}-2\hbar \alpha(t)x\phi
	_{x}-\hbar \alpha(t)\phi -\frac{1}{2}m\Omega (t)x^{2}\phi =0 ,
\end{equation}
which we now have to compare with the general version of the target TDSE (\ref{targetTDSE}) obtained from the point transformation (\ref{gammap}). Demanding these equations to be identical leads to the constraints
\begin{equation}
	\frac{\tau _{t}}{\chi _{x}^{2}}=1,~~~B_{0}(x,t)=-2\hbar \alpha  
	(t)x,~~~V_{0}(x,t)=\frac{1}{2}m\Omega (t)x^{2}+\hbar \alpha   (t),
\end{equation}
which we solve to
\begin{eqnarray}
	\tau (t) \!\!&=&\!\!\int^{t}\frac{ds}{\sigma ^{2}(s)}, \\
	\chi
	(x,t) \!\!&=&\!\!\frac{x+\gamma (t)}{\sigma (t)},  \\
	A(x,t\!\!&=&\!\!\exp \left\{ \frac{im}{\hbar }\left[ \left( \gamma _{t}-\gamma \frac{%
		\sigma _{t}}{\sigma }\right) tx+\left( it\   -\frac{\sigma _{t}}{2\sigma }%
	\right) x^{2}+\delta (t)\right] \right\} , \\
	\delta (t)\!\!&=&\!\!\frac{\gamma }{2\sigma }\left( \sigma \gamma _{t}-\gamma \sigma
	_{t}\right) -\frac{i\hbar }{2m}\log \sigma .
\end{eqnarray}
The newly introduce function $\sigma$ has to satisfy the two standard Ermakov-Pinney equations
\begin{equation}
	\sigma _{tt}-\kappa (t) \sigma -%
	\frac{\omega ^{2}}{\sigma ^{3}} =0 \qquad \text{with}\,\, \kappa (t):=\frac{\gamma _{tt}}{\gamma }=2i\alpha   _{t}-4\alpha   ^{2}-\Omega .
\end{equation}
This equations is analytically solved by \cite{Pinney}
\begin{equation}
	\sigma (t)=\left( Au^{2}+Bv^{2}+2Cuv\right) ^{1/2}  \label{solPinney}
\end{equation}
with $u(t)$, $v(t)$ being solutions to $\ddot{u}$ $+\kappa(t)u=0$, $\ddot{v}+\kappa (t)v=0$, and the constants $A,B,C$ restricted as $C^{2}=AB-\omega ^{2}/(u\dot{v}-v\dot{u})$. Having now completely determined the point transformation $\Gamma$, we know how it acts on $\chi$ and $P$. We can then act with it only on $ H_0(\chi)$ and interpret the result as the invariant for $H_s$, i.e. $\Gamma: H_0(\chi) \rightarrow I_{H_{S}}(x,t) $, obtaining after a lengthy calculation
\begin{eqnarray}
	I_{H_S}\!\!\! &=&\!\!\!\frac{\sigma ^{2}}{2m}p^{2}+m\left( \frac{\gamma \omega ^{2}}{%
		\sigma ^{2}}+2i\alpha (\sigma ^{2}\gamma _{t}-\gamma \sigma \sigma
	_{t})-\sigma \sigma _{t}\gamma _{t}+\gamma \sigma _{t}^{2}\right) x +\frac{1}{2}\sigma \left[ 2i\alpha \sigma -\sigma _{t}\right] \{x,p\}\\
	&& \!\!\!\!\!\!\!\!\!\!\!\!+ \frac{m}{2}\left[ \left( \sigma _{t}-2i\alpha \sigma \right) {}^{2}+\frac{%
		\omega ^{2}}{\sigma ^{2}}\right] ~x^{2}+\frac{m}{2}\left( \frac{\gamma ^{2}\omega ^{2}}{\sigma ^{2}}+\gamma
	^{2}\sigma _{t}^{2}+\sigma ^{2}\gamma _{t}^{2}-2\gamma \gamma _{t}\sigma
\sigma _{t}\right) +\sigma
\left( \sigma \gamma _{t}-\gamma \sigma _{t}\right) p \notag . 
\end{eqnarray}
We convince ourselves that $I_{H_S}$ does indeed satisfy the conservation equation (\ref{LRinvdef}) for $H_S$. Thus we have obtained an invariant by a direct calculation and avoided any guesswork of making an Ansatz for the invariants. For a given target Hamiltonian this is traded to making a suitable guess for the reference Hamiltonian.   

We can now embark on the next step in the procedure and construct the Dyson map from the quasi-Hermiticity relation for the invariants (\ref{invquasi}). It turns out that in this case, i.e. for time-independent mass, the Dyson map also has to be time-independent. This can be overcome by introducing a time-dependence into the mass via the parametrisation
\begin{equation}
	 M(t)= m \sigma^{-2s -r}(t),\qquad   s,t \in \mathbb{N} .
\end{equation}  
In this case we may convince ourselves that the $\eta$ calculated in \cite{fringmoussa2} does indeed produce a Hermitian invariant. By utilizing (\ref{invquasi}) we can also compute the new Dyson map
\begin{equation}
	\eta(t) =\exp \left( -\alpha _{R}m\sigma ^{-r-2s}x^{2}\right) , \label{newDyson}
\end{equation}
simply by demanding the result of the adjoint action of $\eta(t)$ on $I_{H_S}$ to be Hermitian. Having obtained $\eta(t)$ we can calculate from the TDDE (\ref{TDDE}) directly the corresponding Hermitian Hamiltonian 
\begin{equation}
	h=\frac{\sigma ^{r+2s}}{2m}p^{2}+\left( 2m\alpha _{R}^{2}\sigma ^{-r-2s}+%
	\frac{1}{2}m\sigma ^{-r-2s}\Omega ^{2}\right) x^{2}+\frac{1}{4}\partial
	_{t}\ln \left( \frac{\sigma ^{r+2s}}{\alpha _{R}}\right) \{x,p\}.
\end{equation}
The special choice $\alpha _{R}=\sigma ^{r+2s}$ implies that $\alpha _{I}=0$ so that the 
coefficient function $\alpha (t)$ becomes real, the Dyson map becomes
time-independent and $h$ reduces to the time-dependent harmonic oscillator. One may of course carry out similar calculations for different choices of the reference and target Hamiltonians, see \cite{fring2021exactly} for more details and examples \\
\underline{Semi-exact solutions from perturbation theory}\\
We have focussed here mainly on the construction of the exact invariants and the aspect of how they can be utilised to obtain the Dyson map and the metric operator. Ultimately one wishes of course to determine also the wavefunctions and therefore the full solution to the TDSE. In principle, with the knowledge of the invariants these can be obtained from solving the eigenvalue equation (\ref{LR1}). This might, however, not be possible in an exact manner. In \cite{BeckyAnd1} we discussed how to use time-independent perturbation theory or the WKB approximation to solve the invariant eigenvalue equations and then how to proceed in finding approximate solutions to the TDSE. By comparing with several exact solutions in some models, we demonstrated that in certain parameter regimes the quality of the approximated solutions is of a rather good quality. 

\subsubsection{Time-dependent Darboux transformations for non Hermitian Hamiltonian systems}
In section \ref{Darbouxsec} we have seen how Darboux transformations can be used to obtain isospectral partner Hamiltonians and thus potentially associate a Hermitian to a non-Hermitian Hamiltonian. In addition, it is often easier to solve the TDSE for one of the partner Hamiltonians than the other, Darboux transformation can facilitate the calculation of the eigenfunctions for the more complicated system. Moreover, the iteration of Darboux transformations enables the construction of multi-soliton solutions in integrable systems, see \cite{matveevdarboux,correa2016regularized,CenFringHir,cen2020nonlocal,CCFsineG}. Given their usefulness, let us now discuss their time-dependent versions. For time-dependent Hermitian Hamiltonian systems this was achieved in \cite{bagrov1} and thereafter generalised to a non-Hermitian setting in \cite{cen2019time}.

We start with the presentation of the intertwining relation for the two Hermitian Hamiltonians $h_{0}$ and $h_{1}$
\begin{equation}
	\ell \left( i\partial _{t}-h_{0}\right) =\left( i\partial _{t}-h_{1}\right)
	\ell ,  \label{HI}
\end{equation}
of the form $h_{j}\left( x,t\right) =p^{2}+v_{j}\left( x,t\right)$, with explicitly time-dependent potentials $v_{j}\left( x,t\right)$, satisfying the TDSEs  $i\partial _{t}\phi _{j}=h_{j}\phi _{j}$ with $j=0,1$.  The intertwining operator solving (\ref{HI}) can then be constructed as a first order differential operator  
    \begin{equation}
    	\ell \left( x,t\right) =-\ell _{1}\frac{u_{x}}{u} +\ell _{1}  \partial _{x}.  \label{ll}
    \end{equation}
involving a particular solution $u(x,t):=\phi _{0}(x,t)$ and an arbitrary time-dependent function $\ell _{1}(t)$, when the two time-dependent potentials are related as
\begin{equation}
 v_{1}=v_{0}+i\frac{\left( \ell
		_{1}\right) _{t}}{\ell _{1}}+2\left( \frac{u_{x}}{u}\right) ^{2}-2\frac{%
		u_{xx}}{u}.  \label{v1}
\end{equation}
A nontrivial solution to the second system in terms of the first is then obtained as
\begin{equation}
	\phi_{1}=\frac{1}{\ell _{1}u^{\ast }}\int^{x}\left\vert u\right\vert ^{2}dx^{\prime }, \quad \text{with} \,\,   \ell _{1}(t)=\exp \left[ -\int^{t}\func{Im}\left( v_{0}+2\left( \frac{u_{x}}{%
		u}\right) ^{2}-2\frac{u_{xx}}{u}\right) dt^{\prime }\right] .
	\label{ntsol}
\end{equation}
which forces the potential to be real $v_{1}=\func{Re}%
\left( v_{0}+2\left( u_{x}/u\right) ^{2}-2u_{xx}/u\right) $ so that we require a different approach for non-Hermitian systems.

\begin{figure}[h]
	\thispagestyle{empty} \setlength{\unitlength}{1.0cm} 
	\begin{picture}(14.48,9.07)(-2.5,6.5)
		\thicklines
		\put(-0.6,12.0){\LARGE{$H_0  \quad \longleftrightarrow \quad h_0 \quad \longleftrightarrow \quad h_1 \quad \longleftrightarrow \quad H_1$}}
		\put(-0.6,10.0){\LARGE{$I_0^H  \quad \longleftrightarrow \quad I_0^h \quad \longleftrightarrow \quad I_1^h \quad \longleftrightarrow \quad I_1^H$}}	
		\put(-0.6,8.0){\LARGE{$\check{\psi}_0 \quad \longleftrightarrow \quad \check{\phi}_0 \quad \longleftrightarrow \quad \check{\phi}_1 \quad \longleftrightarrow \quad \check{\psi}_1$}}
		\put(-0.6,14.0){\LARGE{${\psi}_0 \quad \longleftrightarrow \quad {\phi}_0 \quad \longleftrightarrow \quad {\phi}_1 \quad \longleftrightarrow \quad {\psi}_1$}}
		\put(-0.3,11.8){\vector(0,-1){1.2}}	
		\put(3.7,11.8){\vector(0,-1){1.2}}
		\put(7.4,11.8){\vector(0,-1){1.2}}
		\put(11.3,11.8){\vector(0,-1){1.2}}
		\put(-0.3,9.7){\vector(0,-1){1.1}}	
		\put(3.7,9.7){\vector(0,-1){1.1}}
		\put(7.4,9.7){\vector(0,-1){1.1}}
		\put(11.3,9.7){\vector(0,-1){1.1}}		
		\put(-0.3,12.5){\vector(0,1){1.1}}	
		\put(3.7,12.5){\vector(0,1){1.1}}
		\put(7.4,12.5){\vector(0,1){1.1}}
		\put(11.3,12.5){\vector(0,1){1.1}}		
		\put(1.4,14.6){\LARGE{$\eta_0 $}}
		\put(1.4,12.6){\LARGE{$\eta_0 $}}
		\put(1.4,9.6){\LARGE{$\eta_0 $}}
		\put(1.4,7.6){\LARGE{$\eta_0 $}}		
		\put(5.3,14.4){\LARGE{$\ell $}}
		\put(5.3,12.4){\LARGE{$\ell $}}
		\put(5.3,9.5){\LARGE{$\ell $}}
		\put(5.3,7.5){\LARGE{$\ell $}}
		\put(9.0,14.6){\LARGE{$\eta_1 $}}
		\put(9.0,12.6){\LARGE{$\eta_1 $}}
		\put(9.0,9.6){\LARGE{$\eta_1 $}}
		\put(9.9,7.6){\LARGE{$\eta_1 $}}		
		\thicklines
		\put(0.14,11.67){\vector(-3,2){0.2}}
		\put(0.14,10.72){\vector(-3,-2){0.2}}
		\put(9.0,11.68){\vector(3,2){0.2}}
		\put(9.0,10.71){\vector(3,-2){0.2}}		
		\qbezier(0.1, 11.7)(4.8, 9.9)(9.0,11.7)
		\qbezier(0.1, 10.7)(4.8, 12.5)(9.0,10.7)
		\put(4.5,11.0){\LARGE{$L$}}		
		\qbezier(11.8, 14.0)(13.1, 11.0)(11.8,8.0)
		\put(12.5,11.0){\LARGE{$\alpha_1 $}}		
		\qbezier(-0.9, 14.0)(-2.1, 11.0)(-0.8,8.0)
		\put(-2.2,11.0){\LARGE{$\alpha_0 $}}	
	\end{picture}
	\caption{Hermitian and non-Hermitian time-dependent Hamiltonians $h_{0}$,$h_{1}$,$H_{0}$,$H_{1}$ with respective solutions $\check{\phi}_{0}$,$\check{\phi}_{1}$,$\check{\psi}_{0}$,$\check{\psi}_{1}$ to their TDSEs with their associated Lewis-Riesenfeld invariants $I_{0}^{h}$,$I_{1}^{h}$,$I_{0}^{H}$,$I_{1}^{H}$  and respective eigenstates $\phi _{0}$,$\phi _{1}$,$\psi _{0}$,$\psi _{1}$
		related by time-dependent Dyson maps $\eta _{0}$,$\eta _{1}$, intertwining operators $\ell $,$L$ and phase factors $\alpha _{0}$,$\alpha _{1}$.} \label{scheme}
\end{figure}
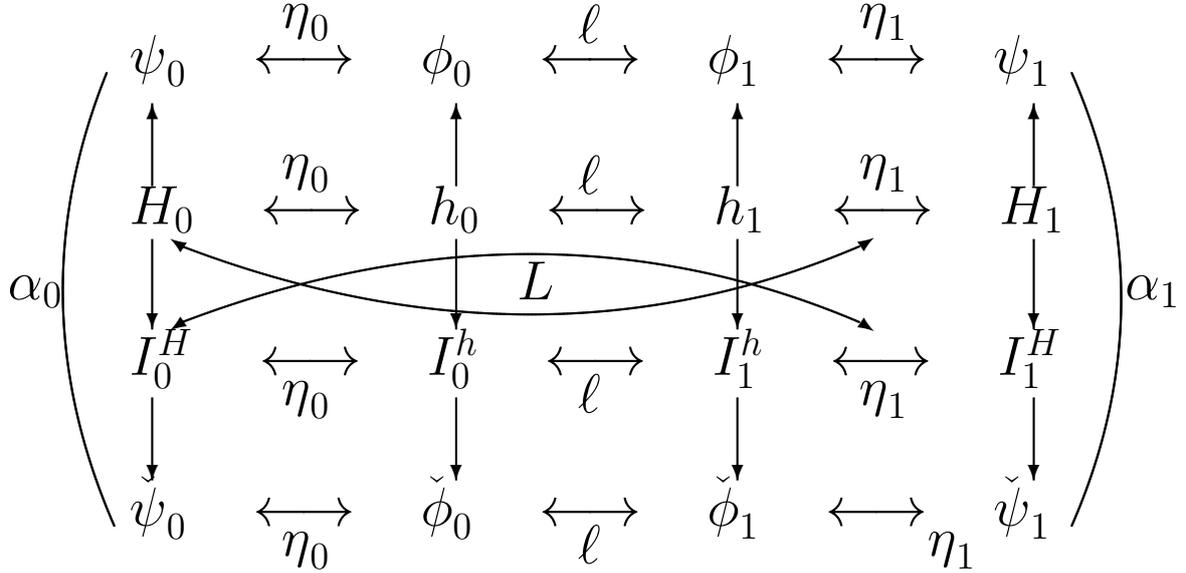

This problem was resolved in \cite{cen2019time}. Considering therefore the two TDSEs $i\partial _{t}\psi _{j}=H_{j}\psi _{j}$ for the non-Hermitian Hamiltonians $H_{0}(t)$, $H_{1}(t)$, we use the TDDE (\ref{TDDE}) for each of them, keep the solutions to each set of equations being related by the Dyson map as in (\ref{phiphi}) and re-write the intertwining relation (\ref{HI}) as
\begin{equation}
	\ell \left( i\partial _{t}-\eta _{0}H_{0}\eta _{0}^{-1}-i\partial _{t}\eta
	_{0}\eta _{0}^{-1}\right) =\left( i\partial _{t}-\eta _{1}H_{1}\eta
	_{1}^{-1}-i\partial _{t}\eta _{1}\eta _{1}^{-1}\right) \ell .  \label{aux}
\end{equation}  
Introducing the new intertwining operator for the non-Hermitian time-dependent Hamiltonians
\begin{equation}
	L:=\eta _{1}^{-1}\ell \eta _{0}.  \label{DarbouxGeneral}
\end{equation}
we can re-arrange (\ref{aux}) into 
\begin{equation}
	L\left( i\partial _{t}-H_{0}\right) =\left( i\partial _{t}-H_{1}\right) L.
	\label{IH}
\end{equation}
 A nontrivial solution to (\ref{IH}) was identified in \cite{cen2019time} as
 \begin{equation}
 	\tilde{\psi}_{1}=  \eta _{1}^{-1}\frac{1}{\ell _{1}\left( \eta _{0}U\right)^{\ast }} \int^{x}\left\vert \eta _{0}U\right\vert ^{2}dx^{\prime } , \label{ntP}
 \end{equation}
where we use the particular solution $\psi _{0}=U=\eta_{0}^{-1}u$. Combining this approach now with the possibility to utilize Lewis-Riesenfeld invariants to construct Dyson maps and eigenfunctions we have a multitude of options at our disposal that we summarize in figure \ref{scheme}. 

		  \subsubsection{Ambiguities and infinite series of Dyson maps}
		It is a well-known feature that the Dyson map, and therefore also the metric, is not uniquely determined by a time-independent Hamiltonian alone \cite{Urubu}. This property can be attributed to certain symmetries of the Hamiltonian \cite{Mostsyme}. With the construction of the Dyson map in (\ref{newDyson}) that differed from the one obtained in \cite{fringmoussa2}, we have already seen that
		also in the time-dependent case the Dyson map is not uniquely pinned down by $H(t)$ alone. We shall demonstrate now that this is due to symmetries of the Lewis-Riesenfeld invariants. Following \cite{fring2021infinite}, we show that one can even construct an infinite series of time-dependent Dyson maps from two different seed maps. 
		
		Thus we start from two different maps  $\eta(t) $ and $\tilde{\eta}(t)$, that  may have been constructed by any of the procedures explained above, obeying two distinct TDDEs
		  \begin{equation}
		  	h=\eta H\eta ^{-1}+i\hbar \partial _{t}\eta \eta ^{-1}\qquad \text{and}%
		  	\qquad \text{ }\tilde{h}=\tilde{\eta}H\tilde{\eta}^{-1}+i\hbar \partial _{t}%
		  	\tilde{\eta}\tilde{\eta}^{-1} . \label{twoTDDE}
		  \end{equation}
		  All three Hamiltonians involved satisfy their own TDSE
		    \begin{equation}
		  		 h(x,t)\phi (x,t)=i\hbar \partial _{t}\phi (x,t), \quad \tilde{h}(x,t) \tilde{\phi} (x,t)=i\hbar \partial _{t}\tilde{\phi} (x,t),  \quad H(x,t)\psi (x,t)=i\hbar \partial _{t}\psi (x,t). \label{twoD}
		  \end{equation}
	We assume further that the two solutions to the TDSE for the Hermitian systems are related to those of the non-Hermitian system by the distinct Dyson maps  	  
		  \begin{equation}
		  	\phi= \eta \psi, \quad \tilde{\phi} = \tilde{\eta} \psi ,
		  \end{equation}
	  which immediately implies that
	  \begin{equation}
	  	\tilde{\phi} = A \phi \qquad \text{with} \quad A:=\tilde{\eta}\eta ^{-1} .
	  \end{equation}
	When eliminating $H$ from the two TDDEs (\ref{twoTDDE}), we obtain an equation of the same form as the TDDEs, but relating two Hermitian Hamiltonians with the Dyson map replace by the newly defined operator $A$
		  \begin{equation}
		  	\tilde{h}=AhA^{-1}+i\hbar \partial _{t}AA^{-1} .
		  \end{equation} 
		The quasi-Hermiticity relation for the invariants (\ref{invquasi}) implies that the respective Lewis-Riesenfeld invariants are related as
		  \begin{equation}
		  	I_{h}=\eta I_{H}\eta ^{-1},~~\ \ I_{\tilde{h}}=\tilde{\eta}I_{H}\tilde{\eta}%
		  	^{-1}, \quad \Rightarrow  I_{\tilde{h}}=A  I_{h} A^{-1} ,
		  \end{equation}
	  where in the last step we eliminated $ I_{H}$ from the first two equations. Using the Hermiticity of the invariants $	I_{h}$ and $I_{\tilde{h}}$ we derive the symmetries of the invariants
		  \begin{equation}
		  		\left[ I_{h},S\right] =0 \quad \text{and} \quad   \left[ I_{\tilde{h}},\tilde{S}\right] =0,
		  		\quad	\text{with} \,\, S:=A^{\dagger }A , \quad \tilde{S}:= A A^{\dagger }. \label{invSym}
		  \end{equation}
	  We can now use this symmetry to generate new Dyson maps in term of the two maps $\eta$ and $\tilde{\eta}$. We find that if and only if $I_{\check{\eta}}=AI_{\tilde{h}}A^{-1}$ is Hermitian then we can define a new Dyson map $\check{\eta}$ associated to a new TDDE 
	  \begin{equation}
	  	 \check{h}=\check{\eta}H\check{\eta}^{-1}+i\hbar \partial _{t}\check{\eta}
	  	\check{\eta}^{-1}, \qquad  \text{with} \,\, \check{\eta}:=\tilde{\eta}\eta ^{-1}\tilde{\eta}, \label{152}
	  \end{equation} 
  involving a new Hermitian Hamiltonian $\check{h}$. 
  
  To prove this let us assume that $\check{\eta}$ is a Dyson map and (\ref{152}) holds. We then compute
  \begin{eqnarray}
  	 \check{h} \!\!&=&\!\!\check{\eta}H\check{\eta}^{-1}+i\hbar \partial _{t}\check{\eta} \check{\eta}^{-1}, \label{153} \\
  	 \!\!&=&\!\! \tilde{\eta} \eta^{-1} \left( \tilde{h} - i \hbar \partial_t \tilde{\eta} \tilde{\eta}^{-1}  \right) \eta \tilde{\eta}^{-1} +i\hbar \partial _{t}\left( \tilde{\eta}\eta ^{-1}\tilde{\eta}  \right)\check{\eta}^{-1} \notag \\ 
  	 \!\!&=&\!\! \tilde{\eta} \eta^{-1}  \tilde{h}  \eta \tilde{\eta}^{-1} 
  	 + i \hbar \partial_t \tilde{\eta} \tilde{\eta}^{-1} - i \hbar \tilde{\eta} \eta^{-1} \partial_t \eta \tilde{\eta}^{-1} \notag \\
  	 \!\!&=&\!\! A\tilde{h} A^{-1}+i\hbar \partial _{t}AA^{-1} ,\notag
  \end{eqnarray}
 which in turn implies
 \begin{equation}
 	I_{\check{\eta}} = A I_{\tilde{h}} A^{-1} = A^2 I_h A^{-2}  . \label{154}
 \end{equation}
Thus if $\check{\eta}$ is a Dyson map then $I_{\check{\eta}}$ is an invariant for the new Hamiltonian $\check{h}$ provided $ A I_{\tilde{h}} A^{-1}$ is Hermitian. In reverse, we can assume $I_{\check{\eta}}$ to be a Hermitian invariant in the form of (\ref{154}) and by inverting all the steps in (\ref{153}) we conclude that $\check{\eta}$ is a Dyson map. 

Similarly, we derive the statement: if and only if $A^{-1}I_{h}A$ is invariant then $\hat{\eta}$ is a new Dyson map with associated TDDEs
	  \begin{equation}
	  	\hat{h}=A^{-1}hA-i\hbar
	  	A^{-1}\partial _{t}A, \qquad \hat{h}=%
	  	\hat{\eta}H\hat{\eta}^{-1}+i\hbar \partial _{t}\hat{\eta}\hat{\eta}^{-1}, \quad \hat{\eta}:=\eta \tilde{\eta}^{-1}\eta
	  \end{equation} 
	 Changing our notation to $\check{\eta}=:\eta _{3}$ and $\hat{\eta}=:\eta _{4}$ we summarize this process as
	  \begin{equation}
	  	\begin{array}{lll}
	  		\eta ,\tilde{\eta} & 
	  		\begin{array}{l}
	  			\nearrow  \\ 
	  			\searrow 
	  		\end{array}
	  		& 
	  		\begin{array}{l}
	  			\eta _{3}=\tilde{\eta}\eta ^{-1}\tilde{\eta}=A\tilde{\eta} \\ 
	  			\\ 
	  			\eta _{4}=\eta \tilde{\eta}^{-1}\eta =A^{-1}\eta 
	  		\end{array}
	  	\end{array}.
	  \end{equation} 
  We can now repeat the above argumentation for different seed Dyson maps $\eta$ and $\tilde{\eta}$ by replacing them with the newly obtained maps $\eta_3$ or $\eta_4$. For instance, we construct two more maps from  
	  \begin{equation}
	  	\begin{array}{lll}
	  		\qquad \qquad	\eta ,\eta _{3} & 
	  		\begin{array}{l}
	  			\nearrow  \\ 
	  			\searrow 
	  		\end{array}
	  		& 
	  		\begin{array}{l}
	  			\eta _{5}=\tilde{\eta}\eta ^{-1}\tilde{\eta}\eta ^{-1}\tilde{\eta}\eta ^{-1}%
	  			\tilde{\eta}=A^{3}\tilde{\eta} \\ 
	  			\\ 
	  			\eta _{6}=\eta \tilde{\eta}^{-1}\eta \tilde{\eta}^{-1}\eta =A^{-2}\eta 
	  		\end{array}%
	  	\end{array}.
	  \end{equation}
  Repeating this process then leads to an infinite series of Dyson maps
  \begin{equation}
  	\begin{array}{lll}
  		\tilde{\eta}^{(n)},\tilde{\eta}^{(m)} & 
  		\begin{array}{l}
  			\nearrow  \\ 
  			\searrow 
  		\end{array}
  		& 
  		\begin{array}{l}
  			\tilde{\eta}^{(2m-n)} \\ 
  			\\ 
  			\tilde{\eta}^{(2n-m)}%
  		\end{array}%
  		,%
  	\end{array}%
  	~~~~~~~~~%
  	\begin{array}{lll}
  		\quad \tilde{\eta}^{(n)},\eta ^{(m)} & 
  		\begin{array}{l}
  			\nearrow  \\ 
  			\searrow 
  		\end{array}
  		& 
  		\begin{array}{l}
  			\eta ^{(2m-n-1)} \\ 
  			\\ 
  			\tilde{\eta}^{(2n-m+1)}%
  		\end{array}%
  		,%
  	\end{array} \label{158}
  \end{equation}
  \begin{equation}
  	\begin{array}{lll}
  		\eta ^{(n)},\tilde{\eta}^{(m)} & 
  		\begin{array}{l}
  			\nearrow  \\ 
  			\searrow 
  		\end{array}
  		& 
  		\begin{array}{l}
  			\tilde{\eta}^{(2m-n+1)} \\ 
  			\\ 
  			\eta ^{(2n-m-1)}%
  		\end{array}%
  		,%
  	\end{array}%
  	~~~~~~~~~%
  	\begin{array}{lll}
  		\eta ^{(n)},\eta ^{(m)} & 
  		\begin{array}{l}
  			\nearrow  \\ 
  			\searrow 
  		\end{array}
  		& 
  		\begin{array}{l}
  			\eta ^{(2m-n)} \\ 
  			\\ 
  			\eta ^{(2n-m)} 
  		\end{array},
  	\end{array}
  \end{equation}
  where
  \begin{equation}
  	\eta ^{(n)}:=A^{n}\eta ,\qquad \tilde{\eta}^{(n)}:=A^{n}\tilde{\eta},~~\ \
  	~\ \ \ \ \text{with }n,m\in \mathbb{Z}   . \label{160}
  \end{equation}
Notice that at each step we need to verify that the new invariants are indeed Hermitian, as otherwise the process breaks down. This possibility may indeed occur.

Let us now return to our example of time-dependent coupled oscillators $H_K(t)$ in (\ref{HKham}). Here we discussed only how wo obtain one Dyson map, but a more systematic study in \cite{fring2021perturb} revealed that more solutions can be found. We take from there the two Dyson maps
\begin{equation}
	\eta= e^{\arcsinh\left(k\sqrt{1+x^2}\right)K_4} e^{-i\arctan(x) K_1}, \qquad
	\tilde{\eta}  = e^{\arcsinh\left(k\sqrt{1+x^2}\right)K_4}e^{i\arctan(x) K_2 },
\end{equation}
as our seed solutions. The combination of Dyson map that relates the two corresponding Hermitian Hamiltonians then results to
\begin{equation}
	A = \tilde{\eta}\eta^{-1} \!\!\! = e^{i \arctan(x)\left(K_1+K_2\right)} .
\end{equation} 
Continuing the iteration procedure as specified in (\ref{158})-(\ref{160}) yields the two infinite series of Dyson maps
\begin{eqnarray}
	\eta^{(n)} \!\! &=& \!\! A^n \tilde{\eta}=e^{\arcsinh(k\sqrt{1+x^2})K_4}e^{i \arctan(x)\left[K_1+(n+1) K_2\right]},\\
	\tilde{\eta}^{(n)} \!\! &=& \!\! A^n \eta= e^{\arcsinh(k\sqrt{1+x^2})K_4}e^{-i \arctan(x)\left[(n+1)K_1+ K_2\right]},
\end{eqnarray}
with corresponding Hamiltonians

\begin{eqnarray} 
	h^{(n)} \!\! &=& \!\!	h^{(1)} + \frac{(n-1)\lambda \sqrt{1+k^2(1+x^2)} }{k (1+x^2)} (K_1+K_2) ,\\
		\tilde{h}^{(n)}\!\! &=& \!\! \left( \frac{\lambda}{2k(1+x^2)} \right) (K_2-K_1)  + \left[ a- \frac{(2n+1)\lambda \sqrt{1+k^2(1+x^2)} }{2k (1+x^2)} \right] (K_1+K_2) , \qquad
\end{eqnarray} 
where
\begin{equation}
		h^{(1)} =  \left[a+\frac{\lambda\left(3\sqrt{1+k^2(1+x^2)}-1\right)}{2k(1+x^2)}\right]\!K_1+\left[a+\frac{\lambda\left(3\sqrt{1+k^2(1+x^2)}+1\right)}{2k(1+x^2)}\right]\!K_2 .
\end{equation}
For this example we can verify at each step that the invariants $I_{	h^{(n)}}$ and $I_{\tilde{h}^{(n)}}$ are indeed Hermitian, thus corresponding to the Lewis-Riesenfeld invariants for the respective Hamiltonians. One may also verify the symmetry relations (\ref{invSym}) for these invariants. We stress that the invariants are not automatically Hermitian as demonstrated explicitly in \cite{fring2021infinite} for a particular choice of the two seed maps. We recall that each of these systems is related to the same common non-Hermitian Hamiltonian $H(t)$, and since the time-dependent metric is different in each case they also lead to different observables, i.e. physics. To make this unique we have to impose additional requirement, such as selecting an additional operator as an observable similarly as in \cite{Urubu} for the time-independent case.  

\section{Applications}  
We conclude our discussion with a few selected applications. The ${\cal PT}$-symmetric regimes are usually nontrivial to identify and tackle when starting from a non-Hermitian Hamiltonian, which is typically simple leading to a complicated, possibly nonlocal, equivalent Hermitian Hamiltonian with an intricate parameter structure. While the details are usually complicated the observed physical phenomena are in general identical to those already observed in Hermitian systems. Novel effects arise at or in the vicinity of the exceptional points, such as the stopping of light \cite{goldzak2018light,miri2019exceptional} or the breakdown of the Higgs mechanism in quantum field theory \cite{alexandre2018spontaneous,mannheim2018goldstone,fring2020goldstone,fring2020t,fring2020pseudo,fring2020massive,fring2021non,takathesis}. The spontaneously broken ${\cal PT}$-regime usually has to be discarded in time-independent quantum mechanical and quantum field theoretical systems, as in this regime we naturally have complex eigenvalues of the energies, or masses, so that one always encounters an infinite grows of energy alongside dissipation. In optical systems, however, one can create simultaneously gain and loss that interestingly mimic this regime. As we have indicated above, also in time-dependent systems the spontaneously broken ${\cal PT}$-regime is mended and becomes physically meaningful. As a particular consequence of this feature we discuss the revival of entropy in section \ref{entropy}.

	\subsection{Optics}
	 Most prominent are applications in optics, which is reflected by the fact that Nature Physics selected ``Parity-Time Symmetry in Optics'' as one of the top 10 physics discoveries between 2005 and 2015 \cite{cham2015top}. The activities trace back to the observation made in \cite{ruschhaupt2005physical} that the Helmholtz equation in the paraxial approximation is formally equivalent to the Schr\"odinger equation involving a ${\cal PT}$-symmetric potential
	 \begin{equation}
	  i \frac{\partial \psi}{\partial z}  + \frac{1}{2k} \frac{\partial^2 \psi}{\partial x^2} + k v(x) \psi =0 ,
	 \end{equation}
	when identifying one of the directions as time $z \rightarrow t$. Here $\psi(x,z)$ is the enveloping function of the electric field $E(x,z)$, $n$ is a refractive index, $n_0$ is a background refractive index, $\omega $ the frequency, $k = n \omega / c$ and  $v(x) = n/n_0 -1$. Refractive indices are complex numbers to that the analogue of the potential is naturally complex and often ${\cal PT}$-symmetric. Since the theoretical identification many experiments have been carried out, see e.g. \cite{PT_optics_ref_1,PT_optics_ref_2,PT_optics_3,PT_optics_4,PT_optics_5},  predicting new phenomena and hence confirming the theoretical formulation as well as the manifestation of $\mathcal{PT}$-symmetric systems in nature. This interesting subject is a vast whole topic in itself and will not be discussed here. We refer the reader to reviews on the subject, such as Jones' chapter 10 in \cite{PTbook} or \cite{el2018non}. 
	
	\subsection{The mended spontaneously broken ${\cal PT}$-regime, entropy revival} \label{entropy}
	As in the explicitly time-dependent systems all regimes, i.e., with intact or spontaneously broken ${\cal PT}$-symmetry and the exceptional point, become physically meaningful  \cite{AndTom3,fring2021perturb}, it is interesting to compare physical quantities across these three regimes.  As an example we consider here the von Neumann entropy. Before discussing an explicit example we set up the relevant equation for the non-Hermitian systems following \cite{fring2019eternal}.

\begin{figure}[h]
	\noindent	\begin{minipage}[b]{0.38\textwidth}     \!\!\!\! \!\!\!\! \includegraphics[width=\textwidth]{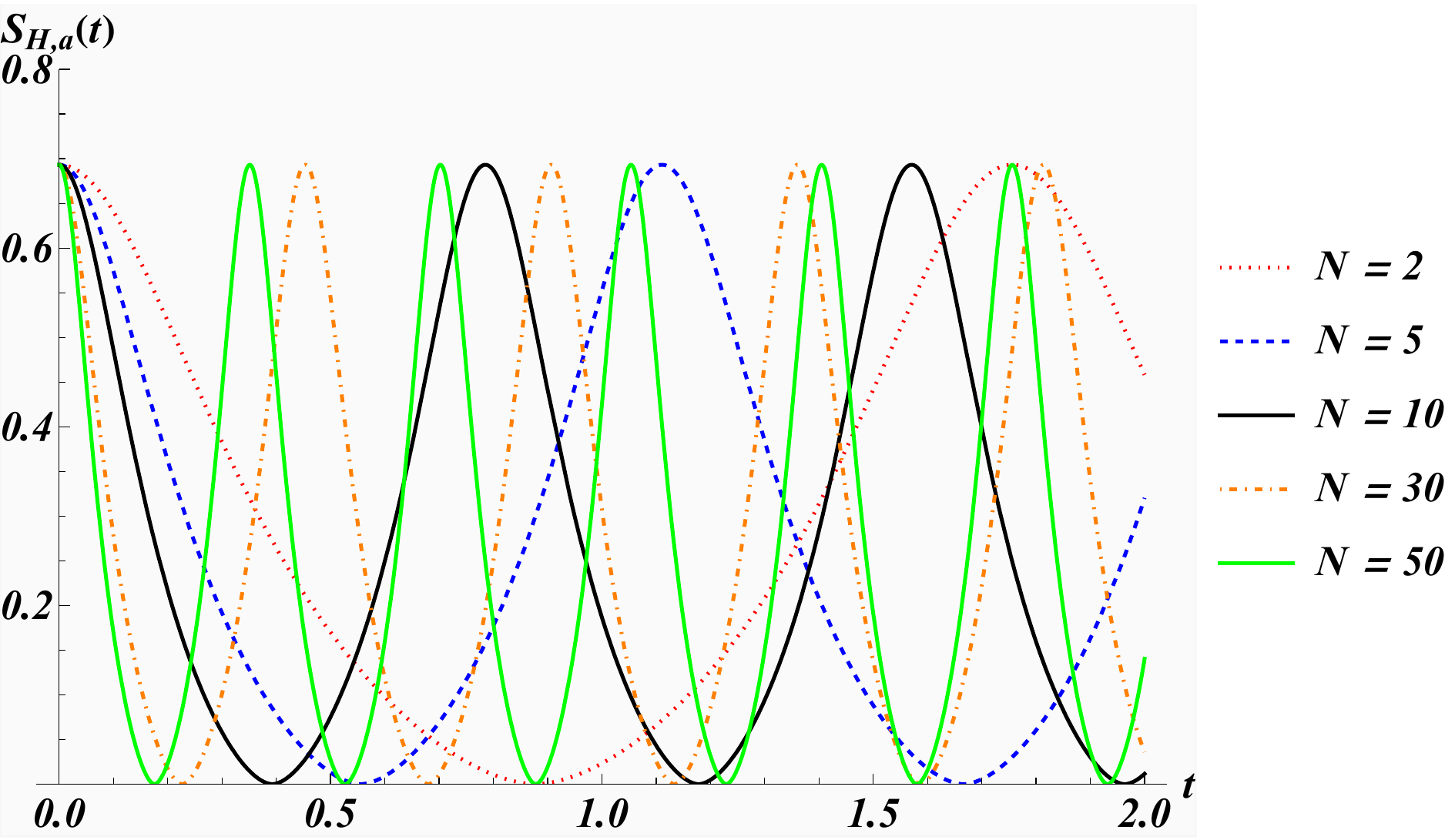}
\end{minipage}
\begin{minipage}[b]{0.31\textwidth}      \includegraphics[width=\textwidth]{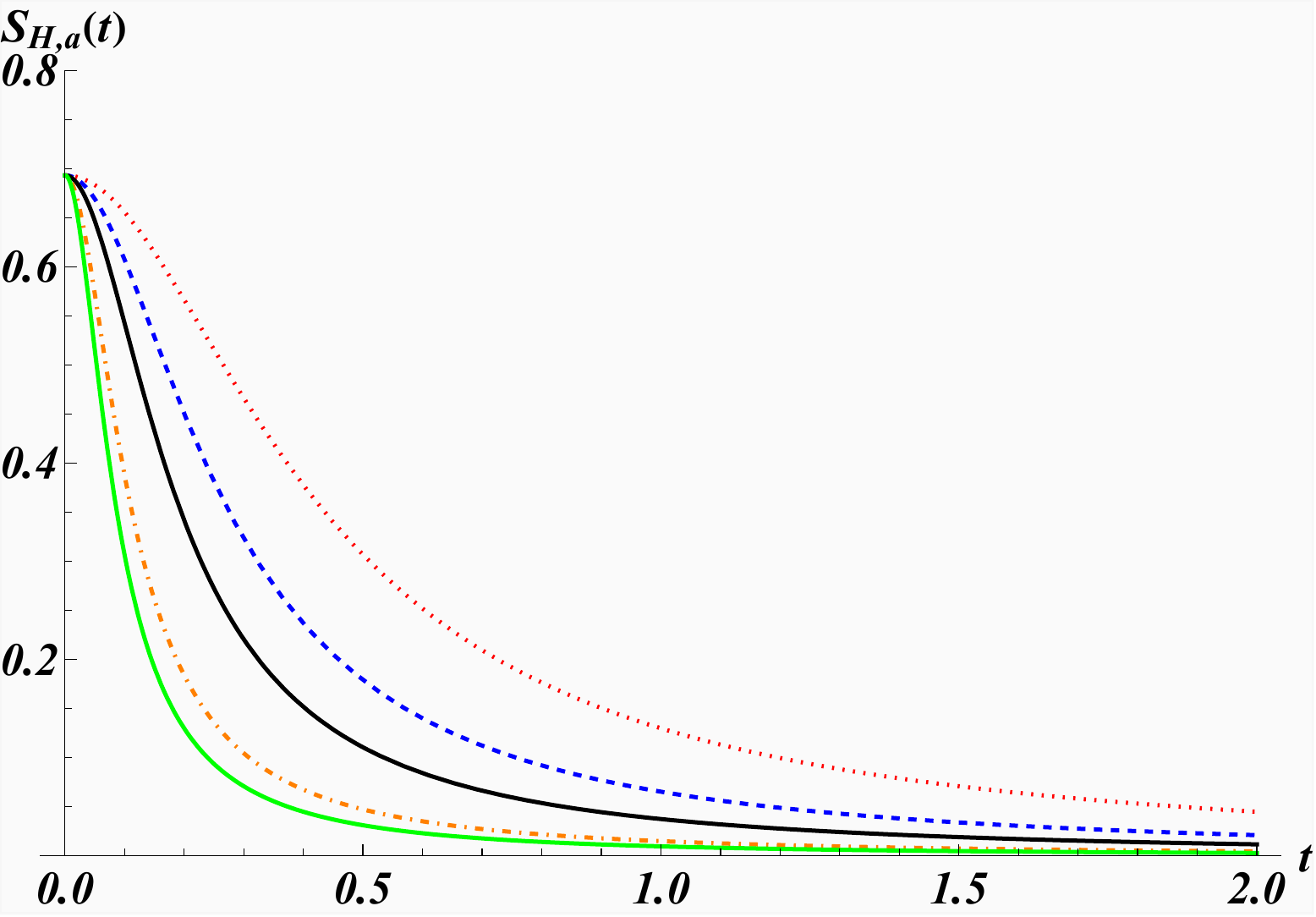}    
\end{minipage}
\begin{minipage}[b]{0.31\textwidth}      \includegraphics[width=\textwidth]{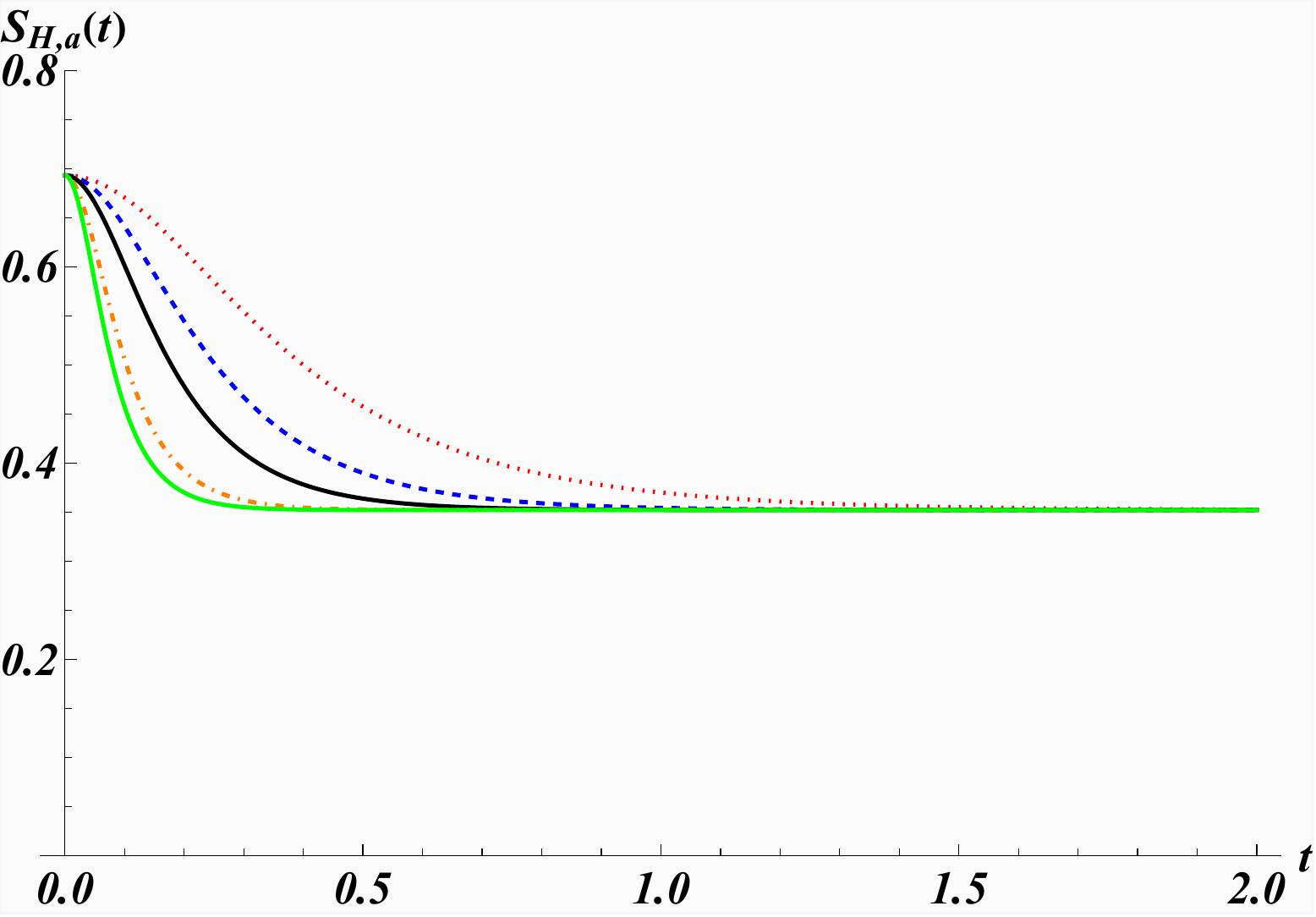}    
\end{minipage}
\caption{Von Neumann entropy of a boson coupled to a bath of different size of $N$ bosons in different $\mathcal{PT}$-regimes. Panel (a): $\mathcal{PT}$-symmetric regime with $c_1=1$,  $\kappa=0.3$, $g=0.7$.  Panel (b): Exceptional point with $c_1=1$,  $\kappa=g$. Panel (c): Spontaneously broken $\mathcal{PT}$-regime with $c_1=1$,  $\kappa=0.7$, $g=0.3$. }
\label{entropy3}
\end{figure}

We start with the definition of the statistical ensemble of states, usually referred to as the density matrix, for a Hermitian Hamiltonian $h$
	\begin{equation}
		\varrho _{h}=\sum\nolimits_{i}p_{i}\left\vert \phi _{i}\right\rangle \left\langle
		\phi _{i}\right\vert .  \label{staten}
	\end{equation}
Here the $p_i$ are the probabilities for the system to be in the pure state $\left\vert \phi _{i}\right\rangle$, satisfying $0 \leq p_i \leq 1$, $\sum_i p_1 =1$. We then consider a system that is composed out of two complementary subsystems $A$ and $B$ with associated eigenstates	$\left\vert n_{i,A}\right\rangle ,\left\vert n_{i,B}\right\rangle $ of $h$. One can then defined a {\em reduced density matrix} for each of the subsystems by means of the following partial traces
	\begin{equation}
		\varrho _{h,A} :=\limfunc{Tr}\nolimits_{B}(\varrho
		_{h})=\sum\nolimits_{i}\left\langle n_{i,B}\right\vert \varrho _{h}\left\vert
		n_{i,B}\right\rangle, \qquad
		\varrho _{h,B} :=\limfunc{Tr}\nolimits_{A}(\varrho
		_{h})=\sum\nolimits_{i}\left\langle n_{i,A}\right\vert \varrho _{h}\left\vert
		n_{i,A}\right\rangle .
	\end{equation}
	The time evolution of the density matrix $\varrho _{h}$ is then governed by Heisenberg's equation of motion
	\begin{equation}
		i \hbar \partial _{t}\varrho _{h}=\left[ h,\varrho _{h}\right] . \label{Heisenbergh}
	\end{equation}
Assuming the density matrices to be quasi-Hermitian
	\begin{equation}
	\varrho _{h}=\eta \varrho _{H}\eta ^{-1}, \label{quasirho}
\end{equation}
and by also using also the TDDE (\ref{TDDE}), we can re-write (\ref{Heisenbergh}) as
\begin{equation}
	i \hbar \eta \left(  \eta^{-1} \dot{\eta} \varrho _{H} + \dot{\varrho} _{H}  -  \varrho _{H} \eta^{-1} \dot{\eta}      \right) \eta^{-1} =
	\eta \left[H,  \varrho _{H} \right]\eta^{-1} + i \bar \eta \left( \eta^{-1} \dot{\eta} \varrho _{H}- \varrho _{H} \eta^{-1} \dot{\eta}      \right) \eta^{-1},
\end{equation}
which implies Heisenberg's equation of motion for the density matrix $\varrho _{H}$ for the non-Hermitian system
	\begin{equation}
		i \hbar \partial _{t}\varrho _{H}=\left[ h,\varrho _{H}\right] .
	\end{equation}
Thus, with (\ref{quasirho}) and the relation between the eigenstates of the Hermitian and non-Hermitian system (\ref{phiphi}), we can re-write (\ref{staten}) as
\begin{equation}
	\eta \varrho _{H} \eta^{-1} = \sum_i p_i \eta \left\vert \psi _{i}\right\rangle \left\langle
	\psi _{i}\right\vert \eta^\dagger , 
\end{equation}
such that the density matrix for the non-Hermitian system becomes
\begin{equation}
	\varrho _{H}=\sum\nolimits_{i}p_{i}\left\vert \psi _{i}\right\rangle \left\langle
	\psi _{i}\right\vert \rho .
\end{equation}
We have now all the ingredients to define the von Neumann entropy the Hermitian and non-Hermitian system
\begin{equation}
	S_{h}=-\limfunc{tr}\left[ \rho _{h}\ln \rho _{h}\right] =-\sum\nolimits_{i}\lambda
	_{i}\ln \lambda _{i}=S_{H}
\end{equation}
The first equality is simply the standard definition of the von Neumann entropy for a Hermitian system. In the second equation we evaluated the trace in terms of the eigenvalues $\lambda_i$ of the density matrix. The equality of $S_{h}$ and $S_{H}$ then simply follows from the quasi-Hermiticity relation for the densities (\ref{quasirho}), which implies that the $\rho$-eigenspectra are identical for both systems. The von Neumann entropies for the subsystems are then defined in a straightforward manner as
\begin{equation}
	S_{h,X}=-\limfunc{tr}\left[ \varrho _{h,X}\ln \varrho _{h,X}\right]
	=-\sum\nolimits_{i}\lambda _{i,X}\ln \lambda _{i,X}=S_{H,X}, \qquad X=A,B . 
\end{equation}
Next we illustrate the working of the above with a simple concrete example described by the Hamiltonian
\begin{equation}
	H_{bb}=\nu a^{\dagger }a+\nu \sum_{n=1}^{N}q_{n}^{\dagger }q_{n}+(g+\kappa
	)a^{\dagger }\sum_{n=1}^{N}q_{n}+(g-\kappa )a\sum_{n=1}^{N}q_{n}^{\dagger }, \qquad \nu, \kappa, g \in \mathbb{R}.
\end{equation}
The model consists of a single boson, created and annihilated by $a^\dagger$, $a$, respectively, coupled to a bath represented by $N$ bosonic fields associated to $q_n^\dagger$, $q_n$, $n=1,\ldots, N$. As one can easily see, $H_{bb}$ is ${\cal PT}$-symmetric with regard to the following antilinear symmetry:  $a\rightarrow-a$, $a^\dagger\rightarrow-a^\dagger$, $q_n\rightarrow-q_n$, $q_n^\dagger\rightarrow-q_n^\dagger$, $i\rightarrow -i$. We will now only discuss the main results and refer the interested reader to \cite{fring2019eternal} for the details of the derivations. Using a standard Fock space representation the system can be characterised by two sequences of infinite eigenstates with energy eigenvalues
\begin{equation}
	E_{m,N}^{\pm }=m\left( \nu \pm \sqrt{N}\sqrt{g^{2}-\kappa ^{2}}\right) ,
\end{equation}	
labelled by $m\in \mathbb{R}$ and depending on the bath size $N$. Evidently the energies are real for $\vert g \vert > \vert \kappa \vert$, what constitutes the $\mathcal{PT}$-symmetric regime, are coalescent at the exceptional point when
$\vert g \vert = \vert \kappa \vert$ and complex conjugate in the spontaneously broken $\mathcal{PT}$-regime for $\vert g \vert < \vert \kappa \vert$. Here our Hamiltonian is time-independent, but we can construct a time-dependent Dyson map according to scenario (ii) in section \ref{scenarios}. The von Neumann entropy for the subsystem consisting of the single boson is then calculated from the partial trace over the bath. In \cite{fring2019eternal} we obtained 
\begin{equation}
	S_{H,a} = -\lambda_- \ln(\lambda_-)  -\lambda_+ \ln(\lambda_+) 
\end{equation}
with
\begin{eqnarray}
		\lambda_\pm \!\! &=&\!\!\left\{ \sin (\gamma ) \sin \left[\mu (t)\right] \pm\cos (\gamma ) \cos \left[ \mu (t) \right] \right\}^2 ,\\
	\mu(t)\!\! &=&\!\!  \frac{1}{2} \arctan \left[\frac{\sqrt{c_1^2+g^2-\kappa ^2} \tan \left(2 \sqrt{\text{N}}  \sqrt{g^2-\kappa
			^2} t\right)}{\sqrt{g^2-\kappa ^2}}\right]  ,
\end{eqnarray}
where $c_1$ is an integration constant. The behaviour of the entropy is qualitatively distinct in the three different $\mathcal{PT}$-regimes as depicted in figure \ref{entropy3}. In the $\mathcal{PT}$-symmetric regime we observe the standard rapid decay of the entropy, also referred to as  ``sudden death'' \cite{yu2009sudden}, which becomes more steep with increasing bath size. At the exceptional point the decay is slightly prolonged and the revival of the entropy observed in the  $\mathcal{PT}$-symmetric regime is absent. Due to the mending of the spontaneously broken $\mathcal{PT}$-regime we can simply continue our analysis into this regime with complex conjugate eigenvalues and notice that the decay is not only further delayed, but even ceases at a finite asymptotic value. Hence this regime may be used to control decoherence, which is of course essential when having applications to quantum computing in mind.

The behaviour presented in this section has also been observed in a non-Hermitian version of the Jaynes–Cummings model \cite{frith2020exotic} and appears to be universal. The features found in the spontaneously broken $\mathcal{PT}$-regime have also been observed in open systems \cite{dey2019controlling}. Evidently more calculations of models and quantities in the spontaneously broken $\mathcal{PT}$-regime are needed. 

\noindent \textbf{Acknowledgments:} I would like to thank Paulo Eduardo Gon\c{c}alves de Assis, Fabio Bagarello, Bijan Bagchi, Olalla Castro-Alvaredo, Andrea Cavagli\`a,  Francisco Correa, Sanjib Dey, Laure Gouba, Boubakeur Khantoul, Thilagarajah Mathanaranjan, Monique Smith, Frederik Scholtz, Takanobu Taira, Miloslav Znojil for discussions and collaboration on the topic of ${\cal{PT}}$-symmetric quantum mechanics and on their time-dependent versions especially Julia Cen, Carla Figueira de Morisson Faria, Miled Hassan Youssef Moussa, Thomas Frith and Rebecca Tenney. 

\section*{References}

\providecommand{\newblock}{}


\end{document}